\documentclass{aa}  
\usepackage{xcolor}

\usepackage{graphicx}
\usepackage{amsmath}
\usepackage{subcaption}
\usepackage{mathtools} 
\usepackage{tabularx}
\usepackage{hyperref}
\usepackage{multirow}
\usepackage{txfonts}
\usepackage{booktabs}
\usepackage[table]{xcolor}
\usepackage[normalem]{ulem}

\begin{document}

   \title{DEMOS} \subtitle{I. A Bayesian inference framework to constrain disc evolution from observed demographics}

   \author{Lorenzo A. Malanga
          \inst{1}\thanks{\email lorenzo.malanga@unimi.it}
          \and
          Giovanni P. Rosotti
          \inst{1}
          \and
          Giuseppe Lodato
          \inst{1}
          \and
          Benoît Tabone
          \inst{2}
          \and
          Alice Somigliana
          \inst{3}
          \and 
          Leonardo Testi
          \inst{4,5}
          }

   \institute{Dipartimento di Fisica 'Aldo Pontremoli', Università degli Studi di Milano, Via Celoria 16, I-20133 Milano, Italy
             \and Université Paris-Saclay, CNRS, Institut d’Astrophysique Spatiale, 91405 Orsay, France
             \and Max-Planck Institute for Astronomy (MPIA), Königstuhl 17, 69117 Heidelberg, Germany
             \and
             Dipartimento di Fisica e Astronomia, Università di Bologna, Via Gobetti 93/2, I-40122 Bologna, Italy.
             \and 
              INAF-Osservatorio Astrofisico di Arcetri, Largo E. Fermi 5, I-50125 Firenze, Italy.
             \\
             }

   \date{Received 16 June 2026; accepted 31 July 2026}

 \abstract{The mechanism driving the evolution and dispersal of protoplanetary discs has a substantial impact on the theories of planet formation. 
 To date, neither the viscous paradigm nor the Magneto-Hydrodynamic (MHD) wind-driven framework have been ruled out by the observations.
 The statistical richness of the observed samples provides a unique opportunity to test evolutionary models through population synthesis.
 However, comparisons between models and observations are normally carried out by eye, and a more statistically rigorous approach has not yet been implemented.
 We present \texttt{DEMOS}, a Bayesian inference tool that aims to infer the model parameters and initial conditions that best reproduce the full distribution of global properties of disc populations.
 This framework relies on a comparison between observations and synthetic populations generated with the 1D disc population-synthesis code \texttt{Diskpop}.
 We tested the tool with both the MHD wind model and the viscous framework with internal photoevaporation.
 Applying \texttt{DEMOS} to synthetic populations, we find that, for both models, it accurately recovers the model parameters of disc populations and their associated uncertainties, including cases in which initial conditions are correlated with the stellar mass. 
 In the optimal scenario in which disc masses, accretion rates and disc sizes are available, the tool can also constrain the angular-momentum transport parameter.
 We show that, although parameter estimates are the most precise with samples of $\sim$ 100 discs, the tool also yields meaningful constraints for smaller populations of 10--30 discs.
 We further find that the predictions obtained with \texttt{DEMOS} closely align with those found by previous works when applied to the same datasets.
 We anticipate that \texttt{DEMOS} will be a valuable tool to infer the initial conditions of observed disc populations, to design observational programmes, and to assess which evolutionary model best reproduces the data.
 This statistical framework can be easily extended to test more evolutionary models, such as late-infall or external photoevaporation models.
}

               
\titlerunning{DEMOS: constraining disc evolution from observed demographics}
\authorrunning{Malanga, L. et al}

\maketitle

\section{Introduction}
\label{introduction}

The large number of exoplanets detected so far\footnote{More than 6000 up to 2026 \url{https://exoplanets.nasa.gov/.}} shows that planet formation is a remarkably common process. 
One of the main challenges currently faced by the community is to explain the statistical diversity and occurrence rates of exoplanets with theoretical models of planet formation.
Various processes, such as the formation, growth, and migration of protoplanets, are deeply affected by the evolution of their host disc \citep{Morbidelli_Raymond2016}.
Therefore, it is paramount to understand the mechanisms that drive the evolution and dispersal of protoplanetary discs.
According to the classic viscous paradigm, the accretion process is induced by the redistribution of angular momentum, which is regulated by turbulence \citep{ShakuraSunyaev1973, LyndenBellPringle1974}.
The origin of turbulence is commonly attributed to magneto-rotational instability (MRI; \citealt{BalbusHawley1991}).
On the other hand, the observed short disc lifetime \citep{Hernandez2007, Fedele2010} is explained by photoevaporative winds launched from the disc surface when heated by energetic radiation from the host star \citep{Hollenbach1994, Clarke2001, Alexander2014}.
However, MRI cannot be triggered over large regions called dead zones \citep{Gammie1996, Bai&Stone_2011, Delage2022}.
This theoretical limitation, together with the observational evidence of low turbulence (see \citealt{Rosotti2023} for a review), has paved the way to the emergence of the magneto-hydrodynamic (MHD) wind driven model, which was initially proposed by \cite{BlandfordPayne1982} and refined by \cite{Ferreira1997, Bai2013} and \cite{Bethune2017}; see also \cite{Lesur2021} for a review.
According to this framework, magnetic winds remove both mass and angular momentum, thus causing both accretion and the dispersal of protoplanetary discs.
This mechanism is well described by 1D evolution models \citep{Suzuki2016, Tabone2022model}.
Despite the significant differences between the predictions of the two models \citep{Tabone2022model, Manara2023} and their remarkable impact on theories of planet formation and migration (e.g., \citealt{Nelson2004, Kimmig2020, Yap2024, Ziampras2025}), so far, none of them has been decisively ruled out by observations.

In the last decades, a new generation of telescopes and interferometers has measured global properties for statistically relevant samples of protoplanetary discs across various star-forming regions (see \citealt{Manara2023} for a review). 
Notably, the Atacama Large Millimeter/submillimeter Array (ALMA) has detected the millimetric continuum emission of numerous sources, allowing us to obtain their dust mass and size \citep{Barenfeld2016, Ansdell2016, Ansdell2017, Bergin2017, Tazzari2017, Andrews2018, Cazzoletti2019, Hendler2020, Sanchis2020, Testi2022, vanTerwisga2022, Carpenter2025}.
Assuming a gas-to-dust ratio of 100, the dust mass is usually used as a proxy to measure the total disc mass, although the conversion presents several uncertainties \citep{Miotello2022}.
Moreover, dust is expected to undergo various processes (such as radial drift) that can decouple it from the gas and cause it not to be a reliable proxy for gas properties \citep{Sellek2020, Toci2021, Zagaria2022, Zormpas2022, Delussu2024}.
Alternatively, both disc masses and radii have been estimated using CO as a proxy \citep{Miotello2016, Ansdell2018, Najita2018, Sanchis2021}; in this case, the conversion requires multiple assumptions on the abundance and optical depth of CO in discs, which do not make the measurements trivial \citep{ Toci2023, Trapman2023}. 
\cite{Trapman2024} showed that CO depletion can be estimated using the N$_2$H$^+$ flux, allowing us to obtain more reliable measurements of the disc mass.
This method was applied in the AGE-PRO large programme \citep{Zhang2025, Trapman2025_AGE_PRO}.
Moreover, recent works \citep{Veronesi2021, Lodato2023, Martire2024, Andrews2024, Pezzotta2025, Longarini2025} showed that disc masses can be measured dynamically from rotation curves. 
This method represents a promising way to probe the total mass contained in discs, because it does not have to rely on model assumptions about the tracers, such as the gas-to-dust ratio, the dust opacity, or the CO abundance.
However, this method can currently only be applied to extended and well-resolved sources.
In addition to ALMA, the Very Large Telescope (VLT)/X-Shooter observations of the Ultraviolet (UV) excess has provided measurements of the mass accretion rate for large samples \citep{Alcala2014, Alcala2017, Manara_2017, Manara2020,  Almendros-Abad2024}.

The statistical robustness of the samples represents an unprecedented opportunity to test disc-evolution models with population synthesis.
\cite{Lodato2017} and \cite{Mulders2017} showed that the viscous model naturally reproduces the observed correlation between disc mass, $M,$ and accretion rate, $\dot{M,}$ when dust flux is used as a proxy \citep{Manara2016}.
However, to reproduce the observed scatter, they had to assume a significant spread in the initial conditions and age.
\cite{Somigliana2022} showed that systematic correlations between initial conditions have to be assumed in order to reproduce the observed correlation between stellar and disc properties, lending credence to \cite{Alexander_Armitage2006} and \cite{Dullemond2006}, whose authors asserted it in the pre-X-Shooter era.
To reproduce disc dispersal, a combination of the viscous paradigm and photoevaporative clearing has been explored \citep{Alecander2006b, Owen2012, Ercolano2014, Somigliana2020}, showing that observed disc properties and disc dispersal can be explained by the model.
On the MHD wind side, \cite{Tabone2022} showed that the $\dot{M}-M$ correlation and the disc dispersal can be reproduced while at the same time assuming a purely wind-driven evolution.
\cite{Zallio2024} expanded the picture by adding a correlation between the initial disc mass, $M_0,$ and the accretion timescale, showing that the correlation and spread can be reproduced within a broad region of the parameter space.
Several studies have identified diagnostics to distinguish between disc evolution models, including disc spreading \citep{Rosotti2019, Trapman2020}, the distribution of accretion rates \citep{Alexander2023}, the spread in $M/\dot{M}$ \citep{Somigliana2023}, the evolution of the slopes of the correlations between disc properties and $M_\star$ \citep{Somigliana2024}, and the evolution of the median disc mass \citep{Malanga2025}.
Nevertheless, none of these results conclusively rule out either of the two models.

The current state-of-the-art of disc population-synthesis model is represented by the work of \cite{Tabone2025}, whose authors developed a systematic method to reproduce the data from the AGE-PRO large programme \citep{Zhang2025}. 
\cite{Tabone2025} used four observables to constrain four free parameters for both the MHD winds and the turbulent-driven model.
The considered observables were the means of the observed, $M$, $\dot{M,}$ and radius, $R,$ distributions and the disc fraction of the Lupus disc population.
Despite the limited sample size, this method provided robust estimates of the free parameters due to the fact that different observables were combined.
The authors found an overall preference for the wind-driven model, as the viscous framework struggled to reproduce discs with low $M$ and high $\dot{M}$.
However, this approach relied on a step-by-step fitting procedure tailored to the adopted disc-evolution models, and it was not able to exploit correlations between parameters or observables or rigorously quantify the uncertainties.
Nevertheless, before ruling out evolutionary models, it is essential to conduct a broad exploration of the parameter space.
This is possible as the observed distributions of disc demographics exhibit spreads and correlations, which allowed us to constrain more than one parameter per observed quantity.
However, a framework that aims to reproduce the full distribution in the observables space using statistically rigorous methods has never been implemented before in the context of protoplanetary-disc populations.

In this study, we present \texttt{DEMOS}, a Bayesian inference tool designed to exploit observed disc demographics with statistically robust techniques to constrain model parameters and initial conditions.
The paper is structured as follows.
In Sect. \ref{methods} we describe the methods used by \texttt{DEMOS} to estimate the likelihood of a set of model parameters given an observed population.
In Sect. \ref{Ruden} we present our generalisation of the \cite{Ruden2004} solution for viscous discs affected by internal photoevaporation and discuss why it is relevant for our work.
In Sect. \ref{tests} we show the tests that we performed in order to verify the validity of the tool (Sect. \ref{test_synthetic_data}) and to understand the impact of different observables and sample sizes (Sects. \ref{test_observables} and \ref{test_size}, respectively) on parameter inference.
In Sect. \ref{comparison} we compare the predictions of our tool to those obtained by previous works; finally, in Sect. \ref{Discussion} we discuss the multiple future applications of \texttt{DEMOS}. 
In Malanga et al. (2026b) (hereafter Paper II) we apply \texttt{DEMOS} to the Lupus demographics and show which initial conditions best reproduce the data for both the viscous and wind-driven frameworks.

\section{Methods}
\label{methods}

 The disc evolution model from observed statistics (\texttt{DEMOS}) is a Bayesian inference framework designed to exploit the statistical richness of observed protoplanetary-disc populations to constrain theoretical model parameters in a rigorous and statistically motivated fashion. 
The tool analyses the distribution of the observed discs within an N-dimensional space, where each axis corresponds to a macroscopical observable, such as disc mass ($M$), mass-accretion rate ($\dot{M}$), disc radius ($R$), and stellar mass ($M_\star$). 
This approach, in contrast to the study of the mean and spread of individual observables (e.g. \citealt{Lodato2017, Tabone2022, Tabone2025}), allows the model to also consider potential correlations among the properties of the young stellar objects (YSOs); this makes it possible to constrain more complex evolutionary models, which need a higher number of free parameters, such as the ones that expect a correlation between initial disc properties and $M_\star$ \citep{Alexander_Armitage2006, Dullemond2006, Somigliana2022, Somigliana2024}.
In addition, \texttt{DEMOS} also takes into account the fraction of surviving discs in the observed star-forming region.

The user can select the evolutionary model that they wish to test. 
This involves specifying values for the parameters that will remain fixed, as well as those that \texttt{DEMOS} will explore. 
We denote the set of free parameters as $\boldsymbol{\theta} = \{ \theta_1. \theta_2, ..., \theta_k\}$ throughout the paper. 
For each free parameter within $\boldsymbol{\theta}$, the user must select the boundaries and the shape of its prior distribution, $p(\boldsymbol{\theta})$.
To assess the compatibility between the data and a model, we need to calculate the probability density function (PDF) of the model in the observational space and compare it to the data.
However, obtaining the PDF directly from the model's $\boldsymbol{\theta}$ is not straightforward.
The model parameters describe the initial conditions of a population, i.e. its status before entering the protoplanetary-disc phase, or theoretical properties that cannot be observed.
Therefore, there is no analytical relation that maps them to the distribution of observables.
As a consequence, we used $\boldsymbol{\theta}$ to generate a synthetic population of discs at a given age, which allowed us to calculate an approximated model PDF in the space of the observables, and we adopt it to evaluate the likelihood of the data. 
We discuss the details of the generation of synthetic populations in Sect. \ref{synthetic_population}. After obtaining a synthetic population, $S$, which identifies the predictions of the model parameters, $\boldsymbol{\theta}$, we evaluated the likelihood of obtaining the data, $D,$ given $S(\boldsymbol{\theta})$. This procedure is described in Sect. \ref{likelihood}.
Eventually, we made use of the Bayes' theorem in its un-normalised form to compute the posterior distribution \begin{equation}
    p(\boldsymbol{\theta}|D) = p(D|\boldsymbol{\theta}) \ p (\boldsymbol{\theta})  = p(\boldsymbol{\theta}) \int p(D|S) \  p(S|\boldsymbol{\theta}) \ \text{d}s = \textit{p(D|S) p}(\boldsymbol{\theta}),
\end{equation} 
where we assume that the prediction of model parameters, $\boldsymbol{\theta,}$ is approximated by the synthetic population, $S$.
Then, we sampled the posterior distribution of $\boldsymbol{\theta}$ with the Markov chain Monte Carlo (MCMC) Python package \texttt{emcee} \citep{Foreman-Mackey2013}. 

\subsection{Synthetic populations}
\label{synthetic_population}

We used the 1D code \texttt{Diskpop} \citep{Somigliana2024} to generate and evolve a population of Class II YSOs, which represents the theoretical prediction of the set of model parameters, $\boldsymbol{\theta}$. 
The zero-age population was obtained via Monte Carlo methods, drawing initial disc conditions from pre-defined distributions, whose parameters are either fixed or within $\boldsymbol{\theta}$. 
The stellar mass, $M_\star,$ of each YSO was extracted from the chosen initial mass function (IMF; e.g. \citealt{Kroupa2001}).
This assumption implies that \texttt{DEMOS} relies on the observed sample being representative of the entire disc population, whose $M_\star$ distribution follows the true IMF.
Subsequently, \texttt{Diskpop} evolves each YSO individually until an age, $t,$ solving the master equation of the evolution of the disc surface density, $\Sigma:$

\begin{align}
    \frac{\partial \Sigma}{\partial t} =  \frac{3}{r} \frac{\partial}{\partial r} \left[ \frac{1}{\Omega r} \frac{\partial }{\partial r} (r^2 \alpha \Sigma c_\text{s}^2)\right] + \frac{3}{2r} \frac{\partial}{\partial r} \left[\frac{\alpha_\text{DW} \Sigma c_\text{s}^2}{\Omega}\right] \notag \\
    - \frac{3 \alpha_\text{DW} \Sigma c_\text{s}^2}{4 (\lambda -1) r^2 \Omega} - \dot{\Sigma}_\text{w} \ ,
    \label{eqn:master}
\end{align}
where $c_\text{s}$ is the speed of sound, $\alpha$ is the \cite{ShakuraSunyaev1973} parameter, $\alpha_\text{DW}$ its equivalent in the MHD-wind model \citep{Tabone2022model}, $\lambda$ the magnetic lever arm, $\Omega$ the rotational angular velocity, and $\dot{\Sigma}_\text{w}$ the photoevaporative mass-loss rate. 
The age of the population can also be included among the free parameters. 
Due to the observational evidence of the disc's lifetime \citep{Hernandez2007, Fedele2010}, we considered simulated discs to be dispersed and no longer evolved them once they met the criteria, this choice was motivated by observations.
In particular, to account for the fraction of observed YSOs that show accretion signatures or infared (IR) excess, we removed synthetic objects where either $M$ or $\dot{M}$ falls below the respective observational threshold from the population.
\texttt{DEMOS} runs \texttt{Diskpop} each time that the likelihood is evaluated; therefore, optimising computational efficiency is crucial.
If analytical solutions of Eq. \eqref{eqn:master} are available, such as the self-similar solutions in the viscous case \citep{LyndenBellPringle1974} or the equivalent for the MHD-wind model \citep{Tabone2022model}, the likelihood evaluation takes about five seconds. However, solutions with internal photoevaporation are normally evolved numerically, which can increase the computational cost by two orders of magnitude, resulting in extremely expensive simulations. Therefore, we implemented an analytical solution for a viscous--photoevaporative scenario (see Sect. \ref{Ruden}).
\begin{figure*}[ht!]
    \centering
    \includegraphics[width=0.311\linewidth]{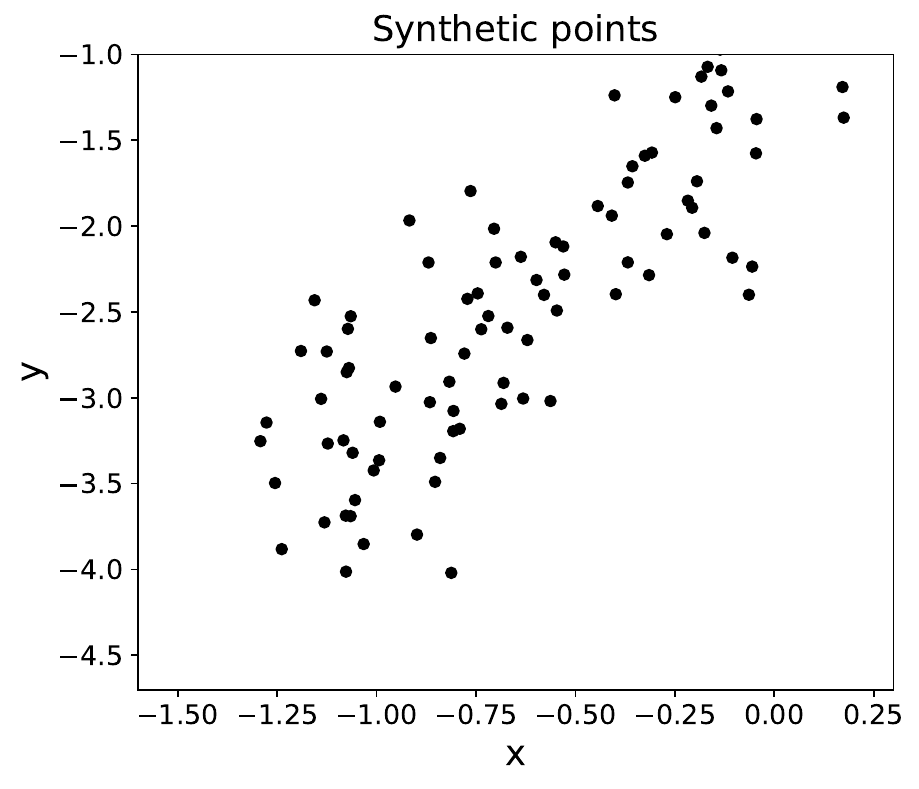} 
    \includegraphics[width=0.33\linewidth]{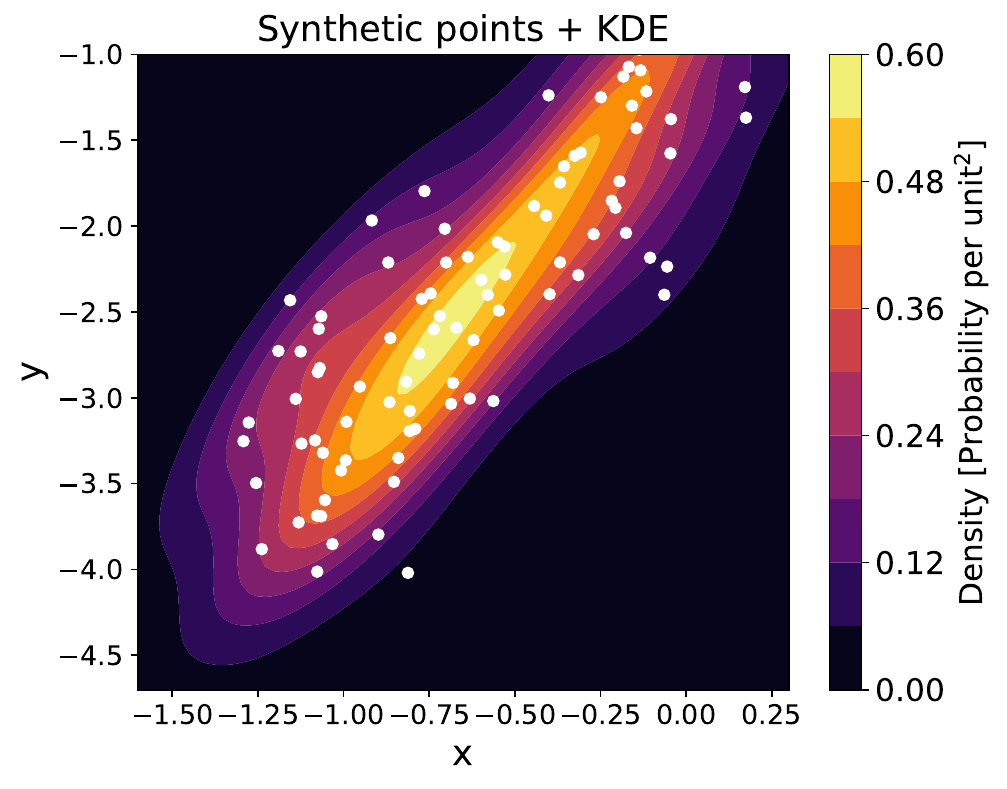} 
    \includegraphics[width=0.33\linewidth]{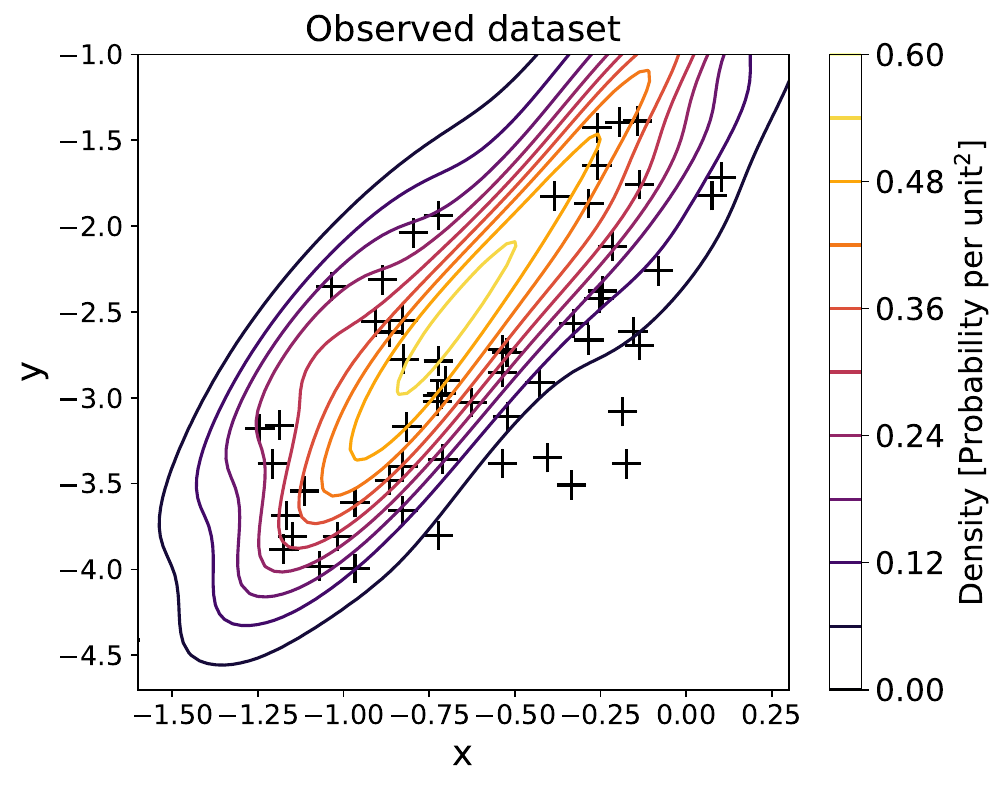} 
    \caption{Illustration of how \texttt{DEMOS} computes the likelihood. For each synthetic point (\textit{left panel}), the KDE is computed, providing the probability density of the model (\textit{middle panel}). Then, the likelihood of obtaining a specific data point in its coordinates is estimated as the value of the KDE in that point (\textit{right panel}). Black (left panel) and white (middle panel) dots represent synthetic objects, while black crosses (right panel) denote observed objects. Contour lines trace regions of equal probability. The process is illustrated in two dimensions, $x$ and $y$, which represent generic observables, but in principle  it is $N$-dimensional.}
    \label{fig:kde}
\end{figure*}
In order to reproduce the intrinsic age scatter of observed disc population \citep{Manara2023}, we introduced a synthetic age scatter of $\sigma_t$ in \texttt{Diskpop}.
Instead of evolving every object until an age of $t$, the code evolves YSOs until an age of $t'_i$, which is drawn from a customisable distribution of $f(\langle t \rangle, \sigma_t)$, where  $\langle t \rangle$ is the mean age.

\subsection{Likelihood estimation}
\label{likelihood}

We computed the likelihood of obtaining the data, $D,$ with $\boldsymbol{\theta}$ given as the compatibility between $D = \{\mathbf{d_i}\}_{i=0}^{N_\text{d}}$ and the synthetic population $S(\boldsymbol{\theta}) = \{\mathbf{s_j}\}_{j=0}^{N_\text{s}}$. Here, $N_\text{d}$ and $N_\text{s}$ represent the number of observed and simulated discs, respectively, and are not required to match. Each $\mathbf{d_i}$ and $\mathbf{s_j}$ is an $N$-dimensional vector in the observable space, where $N$ is the number of observables considered, identifying a single YSO in the corresponding dataset. Therefore, both populations are represented as a distribution of points in the observable space. Since most of the observed populations are approximately log-normal in every dimension, the comparison was carried out in the log--log space.  
For each data point, $\mathbf{d_i}$, we evaluated the likelihood of drawing a point at its specific coordinate given the distribution of synthetic points, which approximates the PDF predicted by the model. The probability density was derived through the kernel distribution estimation (KDE: \citealt{Rosenblatt1956, Parzen1962}). In this approach, a multivariate Gaussian kernel is assigned to each synthetic point,$\mathbf{s_j}$, where the coordinates of the latter are set as the mean of the Gaussian. 
Subsequently, the total density probability function is computed as the average between all the kernels corresponding to the synthetic points $\{\mathbf{s_j}\}$.
The bandwidth matrix, $\Sigma_\text{KDE,}$ follows the Scott rule of thumb \citep{Scott1992}:
\begin{equation}
    \Sigma_{\text{KDE}} \equiv N_\text{s}^{-\frac{1}{N+4}} C \ ,
\end{equation}
where $N$ is the number of dimensions and $C$ the covariance matrix of the synthetic distribution. 
Thus, the per-point likelihood reads
\begin{equation}
    p(\mathbf{d_i}|S(\boldsymbol{\theta})) = \sum_{j = 0}^{N_\text{s}} \frac{1}{N_\text{s} }\mathcal{K} (\mathbf{d_i}| \mathbf{s_j}, \Sigma_\text{KDE}) \ ,
    \label{eqn:likelihood_sum}
\end{equation}
where the kernel, $\mathcal{K} (\mathbf{d_i}| \mathbf{s_j}, \Sigma_\text{KDE}),$ is defined as
\begin{equation}
    \mathcal{K} (\mathbf{d_i}| \mathbf{s_j}, \Sigma_\text{KDE}) = \mathcal{N} (\mathbf{d_i}| \mathbf{s_j}, \Sigma_\text{KDE}) \ ,
    \label{eqn:gaussian_kernel}
\end{equation} 
which is a multivariate Gaussian PDF evaluated at $\mathbf{d_i}$ with mean $\mathbf{s_j}$ and bandwidth $\Sigma_\text{KDE}$, and where $j$ labels synthetic discs.
After evaluating $p(\mathbf{d_i}|S(\boldsymbol{\theta}))$ for each data point, we computed the product to estimate the likelihood of obtaining $D$ given $S(\boldsymbol{\theta})$,  due to the independence of each data point. The relation reads
\begin{equation}
    p(D|S(\boldsymbol{\theta})) = \prod_{i=0}^{N_\text{d}} p(\mathbf{d_i}|S(\boldsymbol{\theta})) \ ,
    \label{eqn:likelihood_prod}
\end{equation}
where $i$ is the index of observed discs.
This procedure is illustrated in Fig. \ref{fig:kde}.
This approach allowed us to assess the compatibility between the synthetic and the observed populations in a comprehensive way, accounting for the mean of the distribution, its shape, its spread, and the correlation between the observables. Consequently, \texttt{DEMOS} succeeds in extracting the maximum amount of information encapsulated in the data, enabling us to constrain a larger number of free parameters than if the observables were evaluated independently (e.g. \citealt{Lodato2017, Tabone2025} among others).

Despite assessing the compatibility between the observed and the synthetic population in the $N$-dimensional observable space, the likelihood term computed from Eqs. \eqref{eqn:likelihood_sum} and \eqref{eqn:likelihood_prod} does not take into account the fraction of surviving discs as a function of age \citep{Hernandez2007, Fedele2010, Ribas2015}. To include this information, we multiplied $p(D|S(\boldsymbol{\theta}))$ by a term that represents the likelihood of observing $N_\text{d}$ survivors out of a total of $N_\text{YSOs}$ in the observed star-forming region, given a corresponding fraction in the synthetic population $f = \frac{N_\text{s}}{N_\text{YSOs, synth}}$.
The total likelihood thus becomes
\begin{equation}
    p(D|S(\boldsymbol{\theta}), N_\text{YSOs}) = p(D|S(\boldsymbol{\theta})) \cdot B(N_\text{d}| N_\text{YSOs}, f) \ ,
    \label{eqn:likelihood_total}
\end{equation}
where 
\begin{equation}
    B(N_\text{d}| N_\text{YSOs}, f) = \binom{N_\text{YSOs}}{N_\text{d}} f ^{N_\text{d}} (1-f)^{N_\text{YSOs}- N_\text{d}}
\end{equation}
is a Binomial function that favours models where the fraction of accreting and disc-bearing YSOs matches the observed one. 
\begin{figure*}[ht!]
    \centering
    \includegraphics[width=0.36\linewidth]{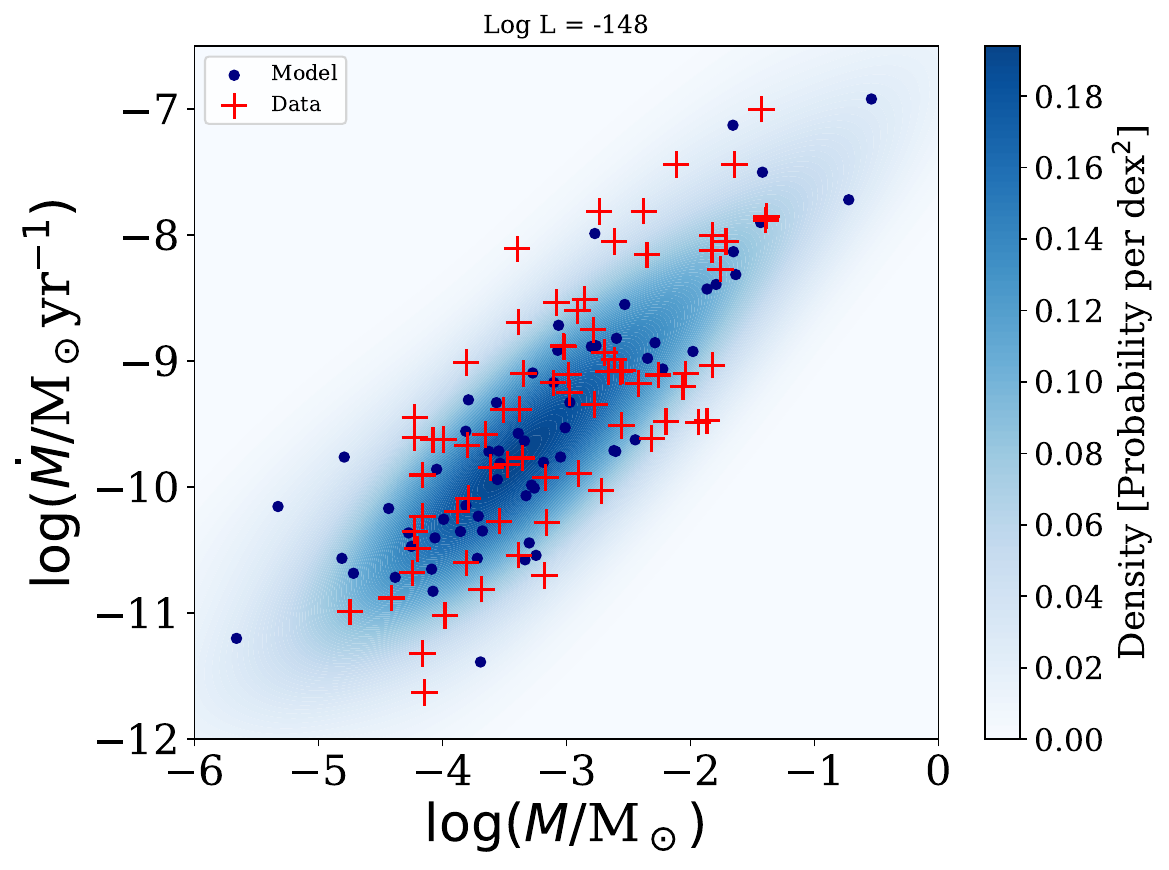} 
    \includegraphics[width=0.36\linewidth]{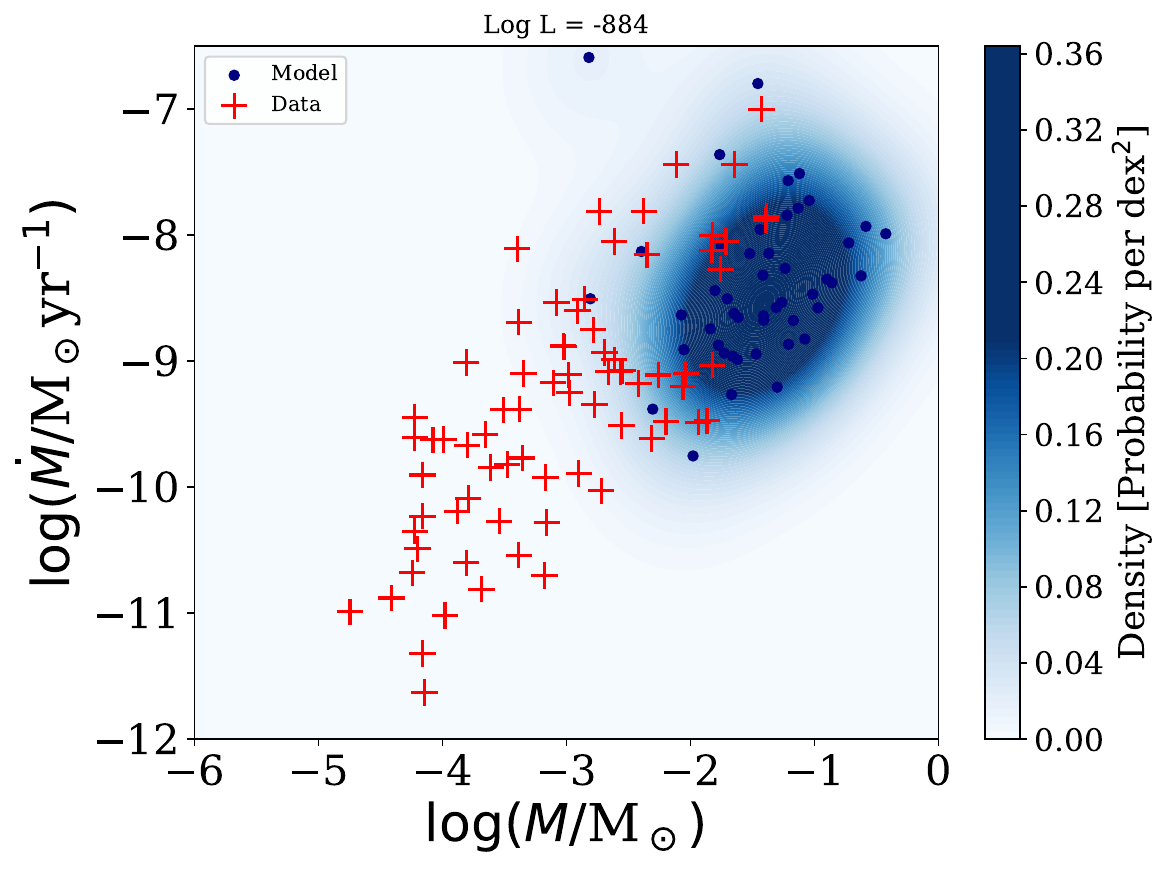} 
    \caption{Comparison between an example observed population (red crosses) and two synthetic populations (blue dots) in the $\log \dot{M} - \log M$ plane. The probability density associated with the model distribution is plotted in blue. In the left panel the two populations closely match, resulting in a high likelihood, $L$, which is defined in Eq. \eqref{eqn:likelihood_prod}. In contrast, the compatibility between the two populations displayed in the right panel is low, and so is the likelihood.}
    \label{fig:data_model_comparison}
\end{figure*}
\texttt{DEMOS} is also suited for analysing multiple populations with different ages. 
In this case, the total likelihood is given by the product of the likelihoods of each individual observed population given the model.

\subsection{Accounting for observational uncertainties}
In the procedure described so far, data points are treated as exact measurements, where their likelihood can be obtained by calculating the probability density from the model at the coordinates of the data (see Eq. \ref{eqn:likelihood_sum}).
However, observed macroscopical properties of discs are always affected by uncertainties, which reflect both instrumental limitations and the robustness of the theoretical assumptions used to convert observed properties to physical ones (see \citealt{Miotello2022} for a review).
In addition, not all measurements in observed samples provide an estimated value; for a subset of sources, only upper limits are available.
As a consequence, we needed to adjust the kernel, $\mathcal{K,}$ of the probability density to deal with these data. 
In general, the $\mathcal{K}$ can be retrieved by convolving the multivariate PDF at $\mathbf{d_i}$ given $\mathbf{s_j}$ and $\Sigma_\text{KDE}$ and an observational kernel, $\mathcal{K}_\text{OBS,}$ which depends on the nature of the observation:
\begin{equation}
    \mathcal{K}(\mathbf{d_i}| \mathbf{s_j}, \Sigma_\text{KDE}, \Sigma_\text{OBS}) = \int \text{d}\mathbf{d'} \ \mathcal{N}(\mathbf{d'}| \mathbf{s_j}, \Sigma_\text{KDE}) \mathcal{K}_\text{OBS}(\mathbf{d'}- \mathbf{d_i}| 0,\Sigma_\text{OBS}) \ ,
\end{equation}
where $\Sigma_\text{OBS}$ is the covariance matrix of observational uncertainties.
In the case of exact detections, we would use 
\begin{equation}
\mathcal{K}_\text{OBS}(\mathbf{t}| \mathbf{d_i}, \Sigma_\text{OBS}) = \delta(t - \mathbf{d_i}) \ ,
\end{equation}
thus retrieving Eq. \eqref{eqn:gaussian_kernel}.
If the observables of a disc are measured with observational uncertainties, their observational kernel is a multivariate PDF with mean of $\mathbf{d_i}$ and covariance matrix of $\Sigma_\text{OBS}$.
Therefore, due to the fact that the convolution between two multivariate distributions is still a multivariate distribution, the kernel becomes
\begin{equation}
    \mathcal{K}(\mathbf{d_i}| \mathbf{s_j}, \Sigma_\text{KDE}, \Sigma_\text{OBS}) = \mathcal{N}(\mathbf{d_i}| \mathbf{s_j}, \Sigma_\text{KDE} + \Sigma_\text{OBS}) \ .
\end{equation}
Accounting for the observational uncertainties can be visualised as a broadening of the KDE distribution of the synthetic population. 
The observational kernel $\mathcal{K}_\text{OBS}$ of sources where only upper limits can be estimated is a multivariate cumulative distribution function (CDF) defined as 
\begin{equation}
    \mathcal{C}(\mathbf{x}| \mathbf{\mu},\Sigma) = \int_{-\infty}^{\mathbf{x}} \text{d}\mathbf{t} \ \mathcal{N}(\mathbf{t} |  \mathbf{\mu}, \Sigma) \ .
\end{equation}
Consequently, the so-called censored kernel reads
\begin{equation}
    \mathcal{K}(\mathbf{d_i}| \mathbf{s_j}, \Sigma_\text{KDE}, \Sigma_\text{OBS}) = \mathcal{C}(\mathbf{d_i}| \mathbf{s_j}, \Sigma_\text{KDE} + \Sigma_\text{OBS}) \ .
\end{equation}
However, in the most general scenario, one source could have $N_{\text{det},i}$ observables that are detected and $N_{\text{upp},i}$ observables where only an upper limit can be estimated, with $N_{\text{det},i} + N_{\text{upp},i} = N$. Therefore, to derive the most suitable kernel, $\mathcal{K,}$ for this type of source, we split coordinate vectors and the covariance matrices in the following way:
\begin{equation}
    \mathbf{d_i} = \binom{\mathbf{d_{i, \text{d}}}}{\mathbf{d_{i, \text{u}}}}  \ ; \ \mathbf{s_j} = \binom{\mathbf{s_{j, \text{d}}}}{\mathbf{s_{j, \text{u}}}} \ ; \ \Sigma = \Sigma_\text{KDE} + \Sigma_\text{OBS} = \binom{\Sigma_{\text{dd} \ \ \ } \ \Sigma_\text{du}}{\Sigma_\text{ud} \ \ \ \Sigma_\text{uu}} \ ,
\end{equation}
where the index $\text{d}$ identifies observables that are detected and $\text{u}$ identifies the ones affected by upper limits. The final expression for $\mathcal{K}$ reads
\begin{equation}
    \mathcal{K}(\mathbf{d_i} | \mathbf{s_j}, \Sigma_\text{KDE}, \Sigma_\text{OBS}) = \mathcal{N}(\mathbf{d_{i, \text{d}}} | \mathbf{s_{j, \text{d}}}, \Sigma_\text{dd}) \cdot \mathcal{C}(\mathbf{d_{i, \text{u}}} | \mathbf{\mu}_{j, \text{u}|\text{d}}, \Sigma_\text{u|d}) \ ,
    \label{eqn:kernel_uncertainties}
\end{equation}
where 
$\mathbf{\mu}_{j, \text{u}|\text{d}} \equiv \mathbf{s}_{j, \text{u}} + \Sigma_\text{ud} \Sigma_\text{dd}^{-1} (\mathbf{d_{i, \text{d}}} - \mathbf{s_{j, \text{d}}})$ and $\Sigma_\text{u|d} \equiv \Sigma_\text{uu} - \Sigma_\text{ud}\Sigma_\text{dd}^{-1}\Sigma_\text{du}$ \citep{Bishop2006}.

The method described in this section represents an optimal way to assess the compatibility between two populations, with higher likelihoods indicating more similar distributions. This is illustrated in Fig. \ref{fig:data_model_comparison}, which compares the same observed population (red crosses) in the $\log M - \log \dot{M}$ plane to two distinct synthetic populations (blue dots) whose PDF is displayed in a blue shade.

\section{Analytical prescription for internal photoevaporation in viscous discs}
\label{Ruden}

\subsection{The analytical solution}

By numerically solving Eq. \eqref{eqn:master}, \texttt{Diskpop} allows us to test a large variety of evolutionary models. However, due to the high computational cost of numerical solutions and the fact that \texttt{DEMOS} needs to evaluate the likelihood many times ($\sim 10^5$) to properly explore the parameter space with \texttt{emcee} or \texttt{UltraNest}, analytical solutions are much more suitable for our purpose. 
Fortunately, an analytical solution for viscous discs affected by internal photoevaporation (i.e. of Eq. \eqref{eqn:master} putting $\alpha_\text{DW} = 0$) exists and was presented and discussed by \cite{Ruden2004}. The solution is, however, limited to a specific photoevaporation profile and requires adaptation to our generic case.
In this section we present the solution obtained by the above-mentioned work, we extend it to any generic profile of mass-loss-rate density, $\dot{\Sigma}_\text{w}(r),$ and we compare it to numerical solutions.
This solution assumes a linear dependence between the kinematic viscosity, $\nu,$ and radius, $r$.

Following \cite{Ruden2004}, we can assume a viscous timescale of $t_{\nu,g} = r_g^2/3 \nu (r_g) $ at the gravitational radius, $r_g = G M_\star/c_\text{s}^2$, and a photoevaporative profile, $\dot{\Sigma}_\text{w}(r)$. 
\cite{Ruden2004} applied the solution only to a specific case where $\dot{\Sigma}_\text{w}(r)$ is a power law, truncated at $r = r_g$; however, we can use an arbitrary profile as long as the integral,
\begin{equation}
    \dot{M}_\text{w} = \int_0^{+\infty} \text{d}r \ 2 \pi r \dot{\Sigma}_\text{w}(r) \ ,
\end{equation}
where $\dot{M}_\text{w}$ is the wind's total mass-loss rate, converges.
We can then re-scale distances and times so that the master equation depends only on the dimensionless quantities $\tau \equiv t/3 t_{\nu,g}$ and $x \equiv (r/r_g)^{1/2}$. 
Assuming $\nu \propto r$, the master equation becomes 
\begin{equation}
    \frac{\partial S}{\partial \tau} = \frac{3}{4} \frac{\partial^2 S}{\partial x^2} - f_m \ ,
    \label{eqn:master_Ruden}
\end{equation}
where $S = \Sigma x^3$ and $f_m = 3 t_{\nu,g} x^3 \dot{\Sigma}_\text{w}$.
Since $r_g$ cancels out in the equation and its value does not affect the result, it can be chosen arbitrarily.
\cite{Ruden2004} showed that it is possible to write an analytic solution of Eq. \eqref{eqn:master_Ruden} in terms of a Green's function $G(x,\tau; x_0, \tau_0)$, known as the `spreading ring' \citep{LyndenBellPringle1974}, which represents the diffusion of an infinitesimally thin disc
\begin{equation}
    G(x,\tau; x_0, \tau_0) = \frac{1}{\sqrt{3 \pi (\tau - \tau_0)}} \left(e^{-(x-x_0)^2/3(\tau-\tau_0)} - e^{-(x+x_0)^2/3(\tau-\tau_0)} \right) \ .
\end{equation}
The formal solution of Eq. \eqref{eqn:master_Ruden} can be written as
\begin{equation}
    S(x,\tau) = S_\text{LBP}(x,\tau)- \Delta S_\text{w}(x,\tau) \ ,
    \label{eqn:analytical_solution_S}
\end{equation}
where $S_\text{LBP}(x,\tau) = \Sigma_\text{LBP}(x,\tau) \cdot x^3$ and $\Sigma_\text{LBP}$ is the self-similar viscous solution, by \cite{LyndenBellPringle1974}, as a function of the dimensionless parameterisations of $r$ and $t$, and 
\begin{equation}
    \Delta S_\text{w}(x,\tau) = \int_0^\tau \text{d}\tau_0 \int_0^{+\infty} \text{d}x_0 \ G(x,\tau; x_0, \tau_0) f_m(x_0, \tau_0) \ .
    \label{eqn:delta_Sw}
\end{equation}
The physical interpretation of Eq. \eqref{eqn:delta_Sw} is that the mass loss due to photoevaporative winds at a given position, $x_0$, and time, $\tau_0$, diffuses through the disc due to viscosity.
Therefore, the surface-density profile at the position $x$ and time $\tau$ is influenced by the mass loss occurring at earlier times, $\tau_0<\tau,$ and at every location, $x_0,$ across the disc. 
As a result, the mass that would otherwise have reached $x$ at $\tau$, but has been instead removed by the winds, is subtracted from the original profile (hence the minus sign in front of $\Delta S_\text{w}$).
In this work, we only considered the wind's mass-loss-rate density, $\dot{\Sigma}_\text{w}$, as a function of radius, assuming it is constant through time. 
As a result, $f_m$ depends only on $x_0$, which allowed us to integrate through time only once, leaving only the integral over $x_0$ to be evaluated for all discs. 
Consequently, instead of numerically evolving Eq. \eqref{eqn:master} until the time, $t$, \texttt{Diskpop} only solves the integral over space, where the time, $t,$ only enters as a parameter. This substantially reduces the computational cost of all \texttt{DEMOS} fits, a Markov chain with 5000 steps, where each walker runs on a dedicated CPU core, requires 3–4 days instead of 15–20 days.

We extended the solution by \cite{Ruden2004} to the $\dot{\Sigma}_\text{w}$ profiles of \cite{Owen2012} and \cite{Picogna2021}, both obtained from hydrodynamical simulations.
The profile of \cite{Owen2012} is a function of $\tilde{x}$, a parameterisation of the radial coordinate defined by
\begin{equation}
    \tilde{x} = 0.85 \left(\frac{r}{\text{au}}\right)\left(\frac{M_\star}{\text{M}_\odot}\right)^{-1} \ .
\end{equation}
This re-scaling implies that the larger the stellar mass, $M_\star$, the farther away the peak of the mass-loss profile, which causes the opening of a gap at larger radii \citep{Clarke2001}.
As a consequence, it is natural to set the parameter, $r_g,$ of the analytic solution to $\tilde{x}$, so that $S(x, \tau)$ is independent of the stellar mass.
Concerning the \cite{Picogna2021} profile, we chose the one computed for a 0.5 M$_\odot$ star and re-scaled it as $\tilde{x}$ = $2 r/M_\star$. 

\subsection{Comparison between analytical and numerical evolution}
\begin{figure}[t]
    \centering

    \begin{subfigure}{0.48\linewidth}
        \centering
        \includegraphics[width=\linewidth]{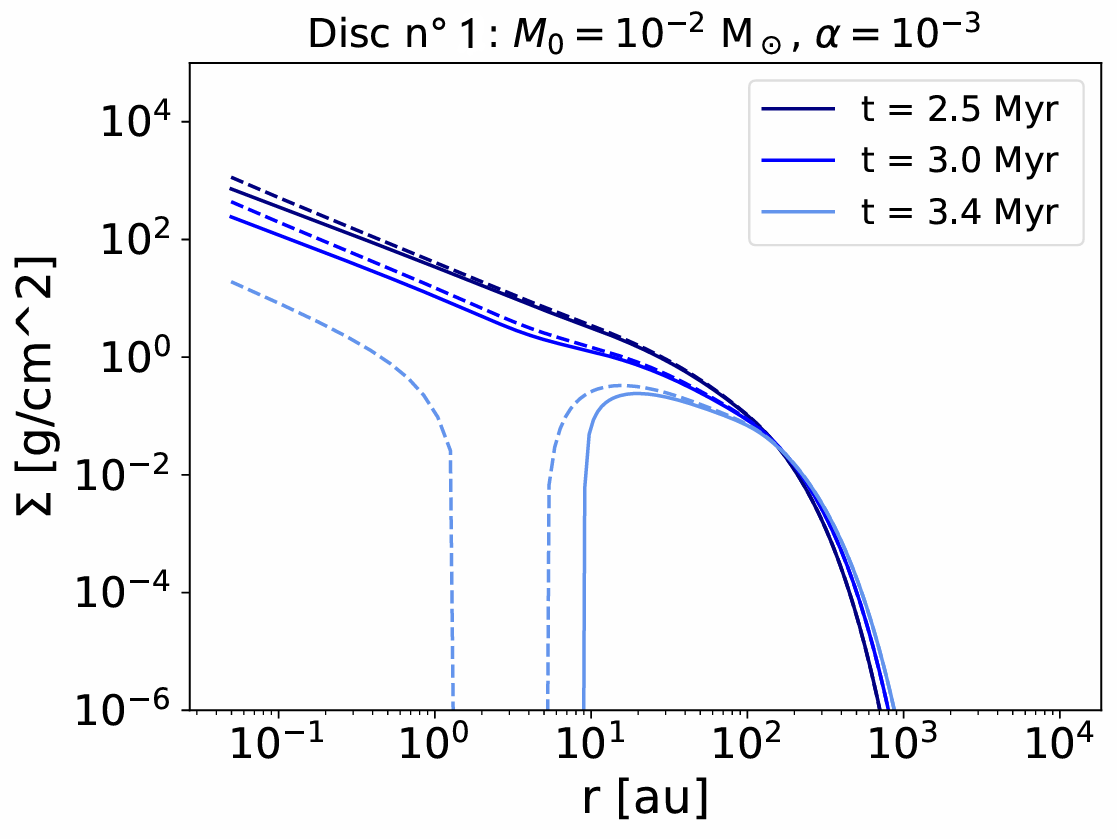}
        \caption{}
    \end{subfigure}
    \hfill
    \begin{subfigure}{0.48\linewidth}
        \centering
        \includegraphics[width=\linewidth]{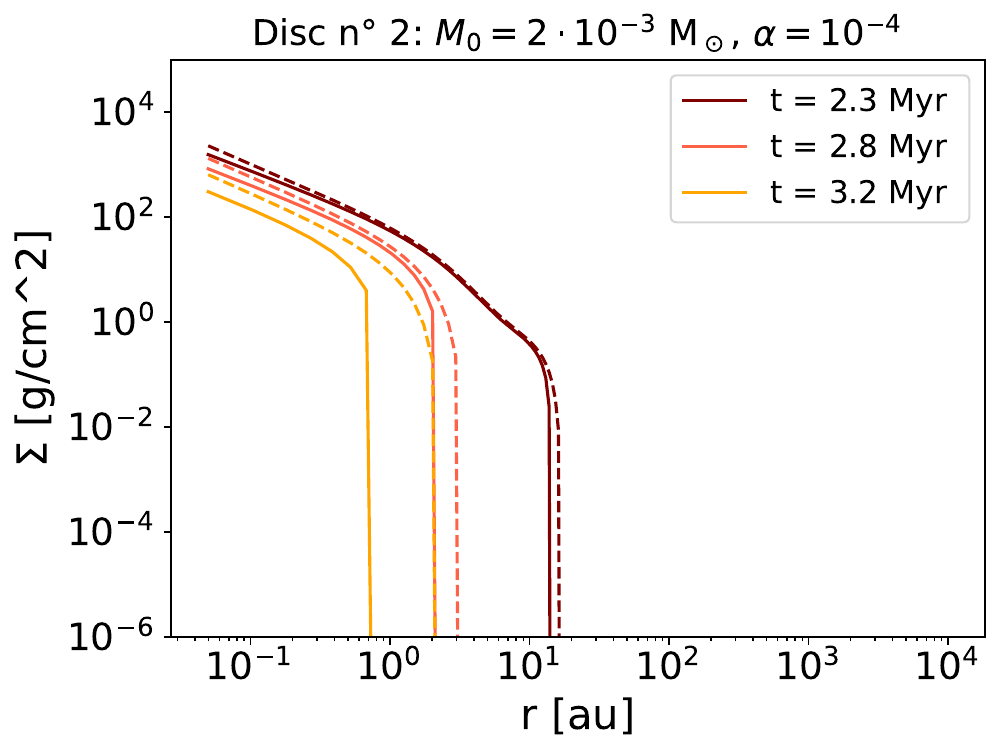}
        \caption{}
    \end{subfigure}

    \vspace{0.05cm}

    \begin{subfigure}{0.47\linewidth}
        \centering
        \includegraphics[width=\linewidth]{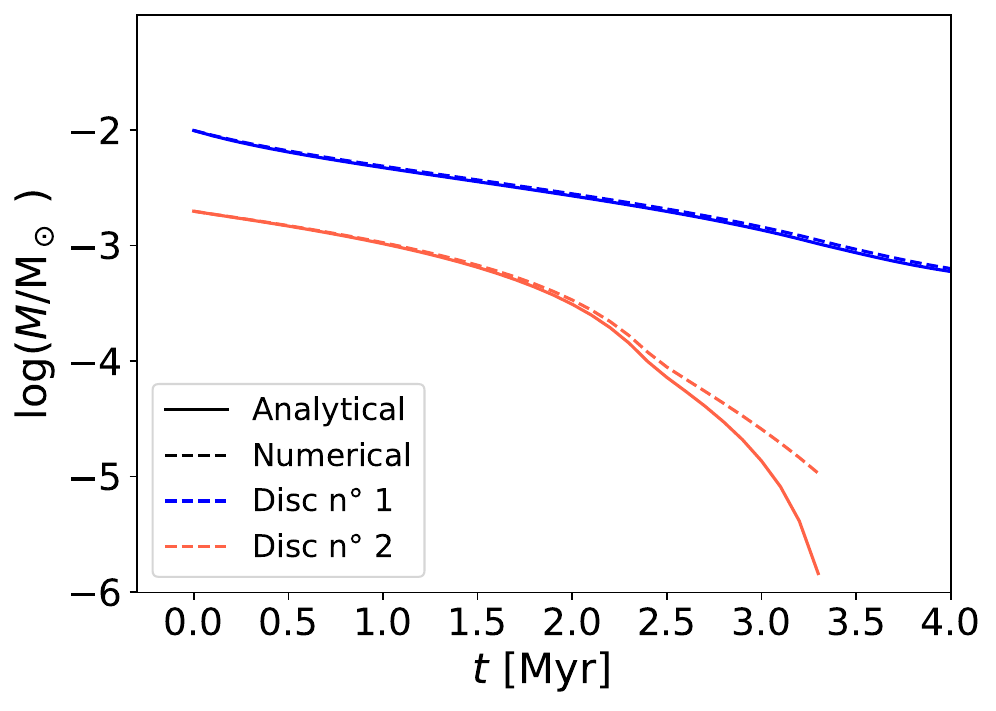}
        \caption{}
    \end{subfigure}
    \hfill
    \begin{subfigure}{0.49\linewidth}
        \centering
        \includegraphics[width=\linewidth]{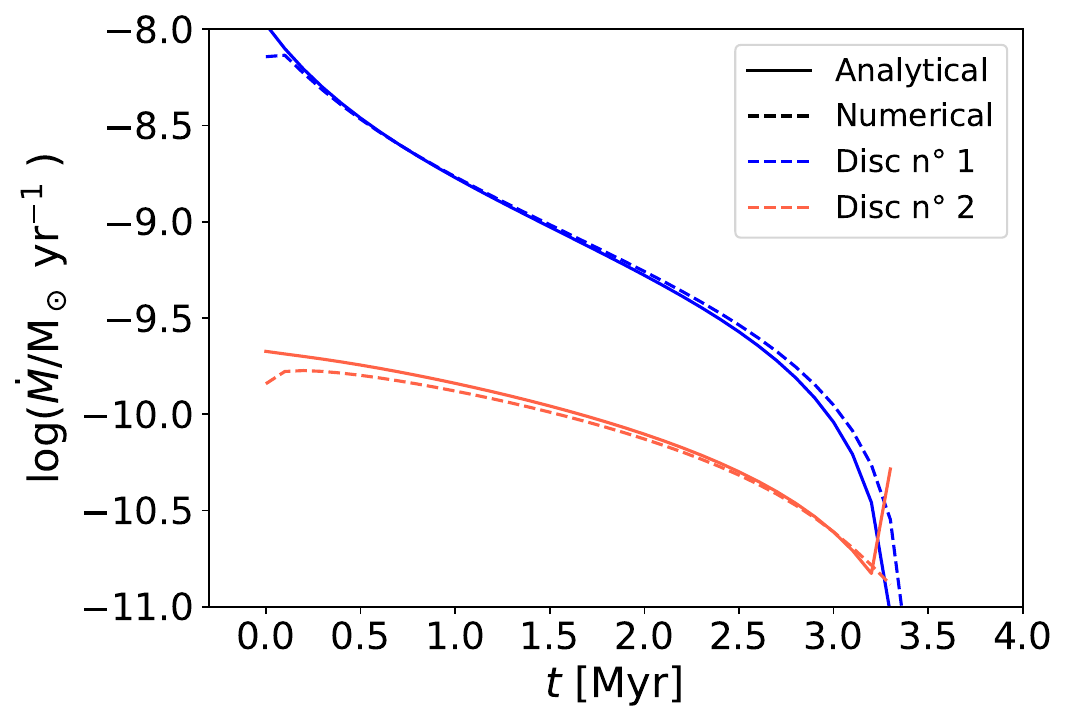}
        \caption{}
    \end{subfigure}

    \caption{Comparison between the analytical (solid line) and numerical (dashed line) evolution of two discs within the viscous--photoevaporative framework. \textit{Top panels}: Surface-density profile, $\Sigma(r),$ for disc n° 1 (\textit{left}), which undergoes I-O dispersal, and n° 2 (\textit{right}), which undergoes O-I dispersal. \textit{Bottom panels}: Time evolution of the disc mass, $M(t)$ (\textit{left}), and the accretion rate, $\dot{M}(t)$ (\textit{right}). For both discs we adopted $R_0 = 10$ au, $\dot{M}_\text{w} = 10^{-9}$ M$_\odot$ yr$^{-1}$, and $M_\star = 1$ M$_\odot$. The first disc has $M_0 = 10^{-2}$ M$_\odot$ and $\alpha = 10^{-3}$, while the second has $M_0 = 2 \cdot 10^{-3}$ M$_\odot$ and $\alpha = 10^{-4}$, corresponding to different disc clearing regimes.}
    
    \label{fig:analytical_numerical}
\end{figure}
A comparison between the analytical and the numerical solution of Eq. \eqref{eqn:master} is necessary to verify that they produce consistent time evolution of the global disc properties, which are the quantities used by \texttt{DEMOS} for likelihood evaluation.
In principle, numerical and analytical solutions of the same equation should give the same result; however, we identified an important difference that arises when the analytical solution \eqref{eqn:analytical_solution_S} vanishes at a certain location, $r_0,$ and time, $t_0$. 
We know from the literature that this scenario results in the opening of a gap in the disc \citep{Clarke2001}.
In numerical solutions, this leads to a decrease of the surface density at all locations of $r<r_0$, because of the absence of material replenishment from $r_0$, where $\Sigma = 0$.
However, in the analytical solution, winds continue to remove material from $r_0$ even at $t> t_0$ despite the presence of the gap, resulting in a much more rapid clearing of the inner disc. 
This occurs because the analytical solution continues to apply mass loss even if the surface density is negative. 
To assess the impact of this discrepancy, we must determine whether the evolution of global disc properties, such as $M$ or $\dot{M}$, differs between the analytical and the numerical solution.  
Therefore, we evolved two synthetic discs with \texttt{Diskpop} using both the analytical and numerical approaches and adopting the $\dot{\Sigma}_\text{w}$ computed by \cite{Owen2012}. 
For both discs we adopted $R_0 = 10$ au, $\dot{M}_\text{w} = 10^{-9}$ M$_\odot$ yr$^{-1}$, and $M_\star = 1$ M$_\odot$, where $R_0$ is the initial truncation radius of the self-similar solution \citep{LyndenBellPringle1974}. 
For disc n° 1 we set $M_0 = 10^{-2}$ M$_\odot$ and $\alpha = 10^{-3}$, while for disc n° 2 we set $M_0 = 2 \cdot 10^{-3}$ M$_\odot$ and $\alpha = 10^{-4}$. 
We selected these sets of parameters so that the two disc have a similar dispersal timescale but follow two different dispersal pathways. 
Indeed, the first disc opens a gap and is cleared from the inside out (I-O), while the second one undergoes an outside-in (O-I) dispersal, a behaviour often obtained in 1D disc-evolution models \citep{Malanga2025, Tabone2025, Picogna2026}.
Figure \ref{fig:analytical_numerical} displays the comparison between the analytical (solid line) and numerical (dashed line) evolution for both disc n° 1 (blue) and disc n° 2 (red).
The top panels shows three snapshots of the surface-density profile, $\Sigma(r)$. For disc n° 1 (top left), the numerical and analytical profiles remain almost identical throughout the disc's lifetime.
As soon as the gap opens at 3.4 Myr, the analytical solution shows a more rapid accretion of the inner disc than the numerical one. 
A similar discrepancy occurs in the evolution of disc n° 2 (top right) at later stages. 
Nevertheless, the analytical and numerical evolutions of $M$ and $\dot{M}$ (in the bottom left and right panels, respectively) show that the discrepancy is negligible for most of the disc's lifetime.
The same argument is valid for the size, $R,$ of both discs.
The discrepancy between the analytical and numerical solutions is also negligible for discs with different properties and lifetimes.
Thus, the \cite{Ruden2004} solution represents a viable method for efficiently generating synthetic populations of discs evolved with the viscous--photoevaporative profile.

\section{Tool validation tests}
\label{tests}

\subsection{Reproduction of a synthetic dataset}
\label{test_synthetic_data}

Before applying the tool to observed data (see Paper II), it is important to test its ability to obtain the posterior distribution of the model parameters of a synthetic dataset, where true values are known.
Consequently, we generated a synthetic population, $D',$ of 200 discs with \texttt{Diskpop}, which we evolved to 2 Myr for both the MHD-wind model and the viscous model with internal photoevaporation. 
During the evolution, a fraction of the discs were dispersed, reducing the size of the synthetic sample to about 100 objects, which served as mock observational dataset for these tests.
For the MHD-wind model we adopted the pure-wind solution of \cite{Tabone2022model}, with $\alpha_\text{DW} \propto (\Sigma_\text{c}(t)/\Sigma_\text{c}(0))^{-\omega}$, where $\omega \in [0,1]$ quantifies the dissipation of the magnetic field and $\Sigma_\text{c}$ is the surface density at the truncation radius.
The parameter $\alpha_\text{DW}$ is inversely proportional to disc magnetisation, $\beta_\text{mag}$ (e.g. \citealt{Armitage2013}).
For the viscous model, we adopted the solution described in Sect. \ref{Ruden}, where $\nu \propto r^\gamma$ and $\gamma = 1$.
For each disc, we set the aspect ratio $H/r$ as
\begin{equation}
    \frac{H}{r} = \frac{1}{30} \left(\frac{M_\star}{\text{M}_\odot}\right)^{-1/2} \left(\frac{r}{ \text{au}}\right)^{1/4}\ ,
\end{equation}
where the scalings correspond to those typically assumed for irradiated discs with a temperature profile of $T \propto r^{-1/2}$.
Then, we applied \texttt{DEMOS} to the synthetic data, assessing whether the posterior distribution, $p(\boldsymbol{\theta}|D'),$ of the model parameters, $\boldsymbol{\theta,}$ is compatible with the ones used to generate $D'$.
For the likelihood evaluation, we followed the procedure described in Sect. \ref{likelihood}, using synthetic populations, $S,$ of 200 discs to estimate the model PDF.
We stress that the size of these synthetic populations is not required to match the number of observed discs, as this does not impact the accuracy of the fit.
Instead, a synthetic population of 200 discs adequately approximates the true model PDF while remaining computationally efficient.
In this test, we considered the stellar mass, $M_\star$; the disc mass, $M$; the mass-accretion rate, $\dot{M}$; the disc radius, $R;$ and the disc fraction as observables. 
We discuss the consequences of considering a different set of observables in Sect. \ref{test_observables}.

For both the viscous and MHD-wind frameworks, we generated a synthetic dataset, $D'$, where $M_\star$  is drawn from the Kroupa initial mass function (IMF; \citealt{Kroupa2001}) for each object.
Let $q$ be a quantity correlated with the stellar mass for every YSO with $M_\star = M_{\star,i}$,
\begin{equation}
    q_i = q' \cdot \left(\frac{M_{\star,i}}{0.3 \ \text{M}_\odot}\right)^{\lambda_q} \cdot 10^{\mathcal{N}(0, \sigma_q)} \ ,
\end{equation}
where $q'$ is the value of that quantity for a YSO with $M_\star$ = 0.3 M$_\odot$ (normalisation), $\lambda_q$ is the slope of the correlation, and $\mathcal{N}$ is a normal distribution with the spread $\sigma_q$.
We included $\lambda_q$ and $q'$ among the free parameters, while we fixed $\sigma_q$ to 0.5 dex.
Table \ref{tab:free_params} shows the quantities that we considered for both models, whether they are correlated with $M_\star$, the free parameters of the MCMC sampling, and the true values.
\begin{table}[h]
\caption{Free parameters used in the MHD-wind and viscous-photoevaporative models.}
\centering
\scriptsize
\renewcommand{\arraystretch}{1.1}
\begin{tabularx}{\columnwidth}{l c X c}
\toprule
\textbf{Quantity} & \textbf{Correlated with $M_\star$?} & \textbf{Parameter} & \textbf{True value} \\
\midrule
\multicolumn{4}{l}{\textbf{Both models}} \\
$M_0$ (M$_\odot$) & Yes & Value for $M_\star$ = 0.3 M$_\odot$ & $2 \cdot 10^{-2}$ \\
 & & Slope of the correlation & 1.8 \\
$\dot{M}_0$ (M$_\odot$ yr$^{-1}$) & Yes & Value for $M_\star$ = 0.3 M$_\odot$ & $6.3 \cdot 10^{-9}$ \\
 & & Slope of the correlation & 1.5 \\
$t$ (Myr) & No & Mean & 2 \\
 & & Scatter & 0.3 \\
\midrule
\multicolumn{4}{l}{\textbf{MHD winds}} \\
$\beta_\mathrm{mag}$ & Yes & Value for $M_\star$ = 0.3 M$_\odot$ & $7.9 \cdot 10^{4}$ \\
 & & Slope of the correlation & $-0.4$ \\
$\omega$ & No & Fixed value & 0.8 \\
$\lambda$ & No & Fixed value & 8 \\
\midrule
\multicolumn{4}{l}{\textbf{Viscous + internal photoevaporation}} \\
$\dot{M}_\mathrm{w}$ (M$_\odot$ yr$^{-1}$) & Yes & Value for $M_\star$ = 0.3 M$_\odot$ & $7.9 \cdot 10^{-10}$ \\
 & & Slope of the correlation & 1 \\
$\alpha$ & No & Fixed value & $10^{-3}$ \\
\bottomrule
\end{tabularx}
\tablefoot{Quantities correlated with $M_\star$ (e.g. $M_0$) have both slopes and values for $M_\star$ = 0.3 M$_\odot$ (normalisation) as free parameters. 
For such correlations, we assumed a spread of 0.5 dex, which is not included among the free parameters.
Quantities kept constant across the population (e.g. $\alpha$) are treated as fixed.}
\label{tab:free_params}
\end{table}
\begin{figure*}[t]
    \centering
    \includegraphics[width=0.7\linewidth]{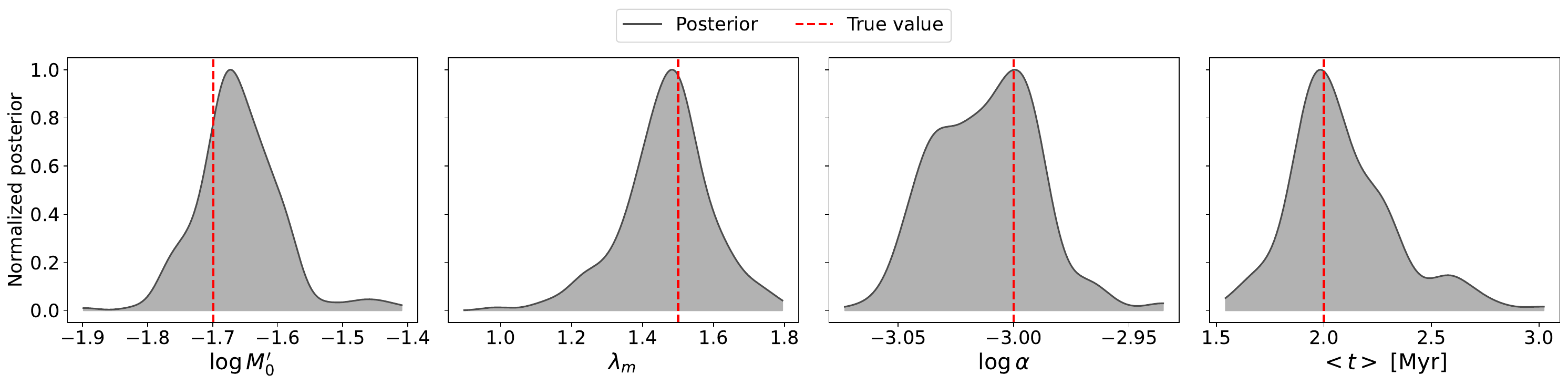} 
    \caption{1D posterior distribution of four of the free parameters of the viscous-photoevaporative model. The parameters are the initial disc mass for a 0.3 M$_\odot$ star, $\log M_0 '$; the slope of the correlation, $\lambda_m$; the $\alpha$ parameter; and the mean age, $\langle t \rangle$. The distribution is plotted in grey, while the vertical dashed red line represents the true value.}
    \label{fig:posterior_viscous}
\end{figure*}
\begin{figure*}[t]
    \centering
    \includegraphics[width=0.9\linewidth]{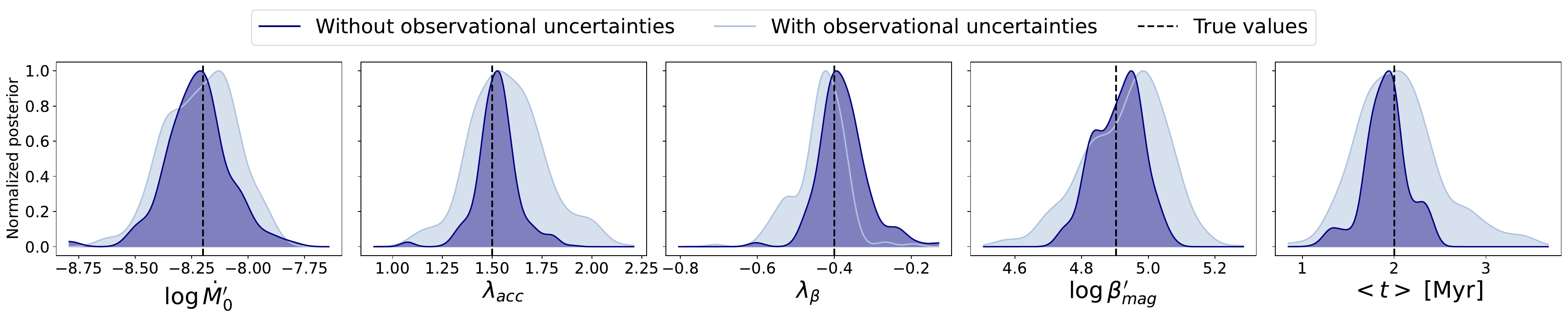} 
    \caption{1D posterior distribution of five of the free parameters of the MHD-wind model with (grey) and without (blue) observational uncertainties. The parameters are the initial accretion rate for a disc around a 0.3 M$_\odot$ star, $\log \dot{M}_0'$; the slope of the correlation, $\lambda_\text{acc}$; the magnetisation parameter for a disc around a 0.3 M$_\odot$ star, $\log \beta_\text{mag}'$; the slope of the correlation, $\lambda_\beta$; and the mean age, $\langle t \rangle$.
    The vertical dashed black line represents the true value.}
    \label{fig:posterior_uncertainties}
\end{figure*}
We included the mean age, $\langle t \rangle,$ and the age scatter, $\sigma_t$, among the free parameters.
For all tests in this section and Sect. \ref{comparison}, we show the resulting posterior distributions obtained with \texttt{emcee}.
Figure \ref{fig:posterior_viscous} shows the 1D posterior distribution of four of the free parameters of the viscous--photoevaporative model: the initial disc mass for a 0.3 M$_\odot$ star, $\log M_0 '$; the slope of the $M_0-M_\star$ correlation, $\lambda_m$; the $\alpha$ parameter; and the mean age, $\langle t \rangle$.
The figure highlights that the input parameter is compatible with the obtained posterior distribution.
Figures \ref{fig:mcmc_mhd_4_variables_corr} and \ref{fig:mcmc_viscous_4_variables_corr} show the corner plots of the posterior distribution of all free parameters for both the MHD-wind model and the viscous model, respectively. The compatibility between the posterior distribution and the true values highlights the correctness of the fitting procedure.

In addition, we performed a fit of the same synthetic populations after adding a scatter to mimic the typical observational uncertainties. We adopted 0.1 dex, 0.5 dex, 0.35 dex, and 0.25 dex for $M_\star$, $M$, $\dot{M,}$ and $R$, respectively, following the criteria described in Paper II.
Figure \ref{fig:posterior_uncertainties} compares the posterior distributions of five of the free parameters of the MHD-wind model, obtained with (grey) and without (blue) observational uncertainties, showing that the inclusion of errors produces an expected broadening of the distributions.
The parameters shown are the initial accretion rate for a disc around a 0.3 M$_\odot$ star, $\log \dot{M}_0'$; the slope of the correlation, $\lambda_\text{acc}$; the magnetisation parameter for a disc around a 0.3 M$_\odot$ star, $\log \beta_\text{mag}'$; the slope of the correlation, $\lambda_\beta;$ and the mean age, $\langle t \rangle$.
Figure \ref{fig:mcmc_viscous_4_variables_corr} compares the posterior distribution of the full set of parameters of the MHD-wind model with and without observational uncertainties.

\subsection{The effect of observables on model-parameter estimation}
\label{test_observables}
\begin{figure}[h]
        \centering
        \includegraphics[
                width=0.96\hsize]{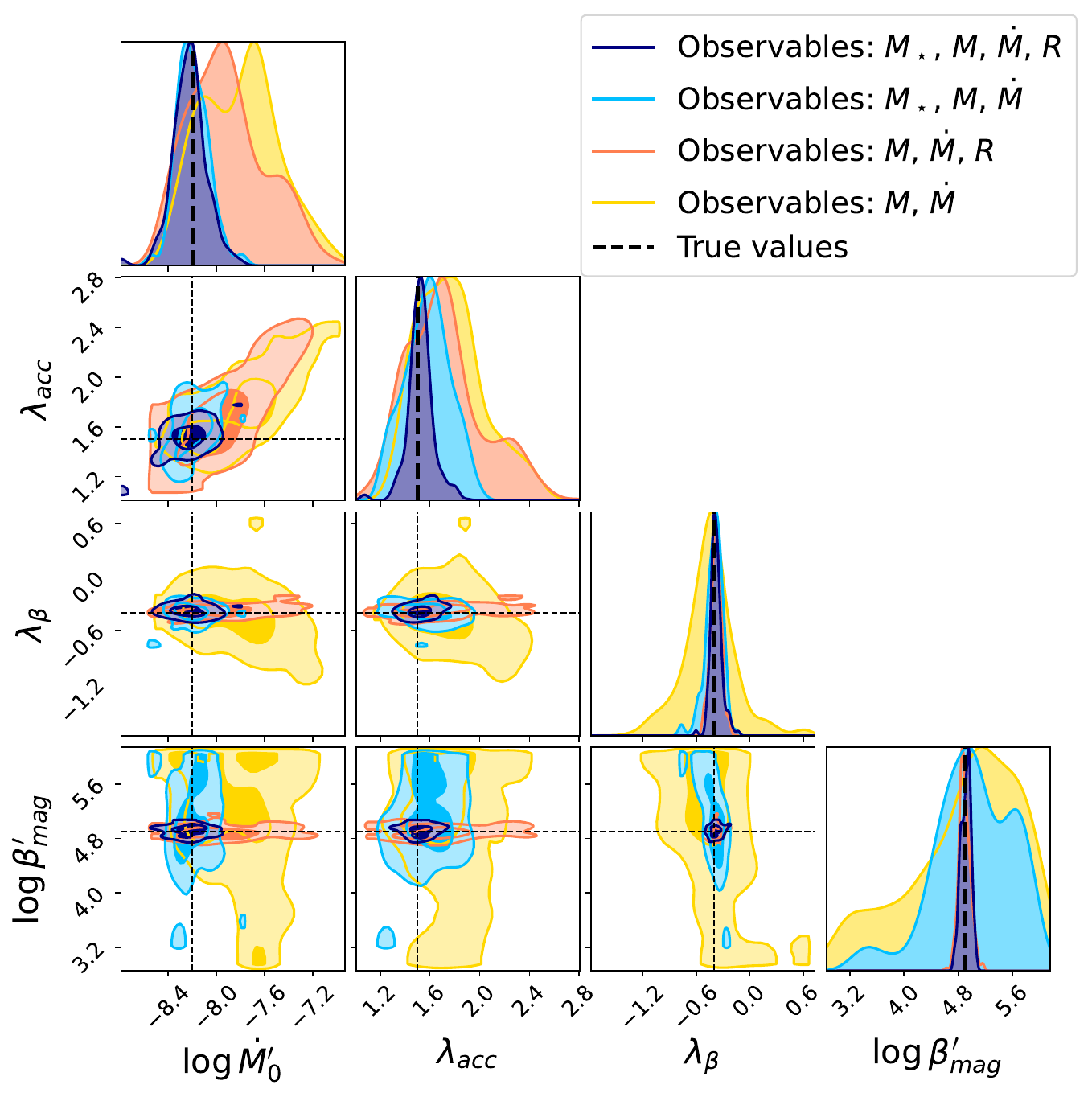}
        \caption{Comparison between the posterior distribution of a subset of the model parameters of a synthetic dataset that evolves according to the MHD-wind model for different sets of observables. The considered observables are $\{M_\star, \ M, \ \dot{M}, R\}$ (blue), $\{M_\star, \ M, \ \dot{M}\}$ (light blue), $\{M, \ \dot{M}, R\}$ (dark orange), and $\{M, \ \dot{M}\}$ (yellow). For all fits, the disc fraction is also included among the observables. The darker and lighter hatched regions include the 1$\sigma$ and 2$\sigma$ confidence intervals, respectively. The dashed black lines show the true values for each parameter.}
        \label{fig:test_observables}
    \end{figure}
Synthetic data can also be used to test which observables are most important to best constrain model parameters.
This test is effective not only to identify the set of free parameters that can be robustly constrained by a given set of observables, but also to understand which quantities need to be measured to infer a particular parameter. 
Thus, we compared the fit for the MHD-wind model presented in Sect. \ref{test_synthetic_data}, which takes into account $M_\star$, $M$, $\dot{M}$, $R,$ and the disc fraction as observables, with fits of the same synthetic population that consider only a subset of the observables.
Figure \ref{fig:test_observables} shows the posterior distribution of four parameters among the set, $\log \dot{M}_0' $, $\lambda_\text{acc}$, $\lambda_\beta,$ and $\log \beta_\text{mag}'$, obtained fitting all four observables (blue), $\{M_\star, \ M, \ \dot{M}\}$ (light blue), $\{ M, \ \dot{M}, \ R\}$ (dark orange), and $\{M, \ \dot{M}\}$ (yellow).
For all fits we include the disc fraction among the observables.
The first two parameters of the plot, $\lambda_\text{acc}$ and $\lambda_{\beta}$, define the correlation between the initial accretion rate and the stellar mass. To optimally constrain their distribution, it is crucial that both $\dot{M}$ and $M_\star$ are included as observables. 
As a consequence, the fits that do not include $M_\star$ (dark orange and yellow) produce a broader posterior distribution for both parameters.
In contrast, $\beta_\text{mag}$ remains completely unconstrained by fits that do not include the radius among the observables. 
This is due to the fact that, by definition, 
\begin{equation}
    \beta_\text{mag} \propto \alpha_\text{DW}^{-1} \propto t_\text{acc,0}\  R_0^{-1}  \ ,
\end{equation}
where $t_\text{acc,0} \propto M_0/\dot{M}_0$ is the typical accretion timescale for MHD-wind-driven models \citep{Tabone2022model}.
While $t_\text{acc,0}$ is constrained by the distributions of $M$ and $\dot{M}$, there is a strong degeneracy between $\beta_\text{mag}$ and $R_0$, which can only be broken by considering $R$ as an observable.
The same argument is valid for $\alpha$ in the viscous model.
As a result, fits that do not include $R$ (light blue and yellow) fail to constrain $\beta_\text{mag}$.
Similarly, the distribution of the slope of the correlation between $\beta_\text{mag}$ and $M_\star$, $\lambda_\beta$, results much wider when neither $R$ nor $M_\star$ are included as observables (yellow).
In general, Fig. \ref{fig:test_observables} highlights that including more observables leads to more accurate and precise posterior distributions of the model parameters.
Moreover, to avoid degeneracies, it is important to select the most suitable set of free parameters based on the available observables in the dataset.
For instance, observed samples that do not include the disc sizes, $R,$ should not be used to constrain the angular-momentum transport parameter $\alpha$ or $\alpha_\text{DW}$; instead, they are better suited to constrain the accretion timescale, $t_\text{acc,0}$.
The distribution of the complete set of parameters is shown by Fig. \ref{fig:test_observables_total}.

\subsection{Effect of the size of the observed sample}
\label{test_size}

\begin{figure}[h]
        \centering
        \includegraphics[
                width=0.96\hsize]{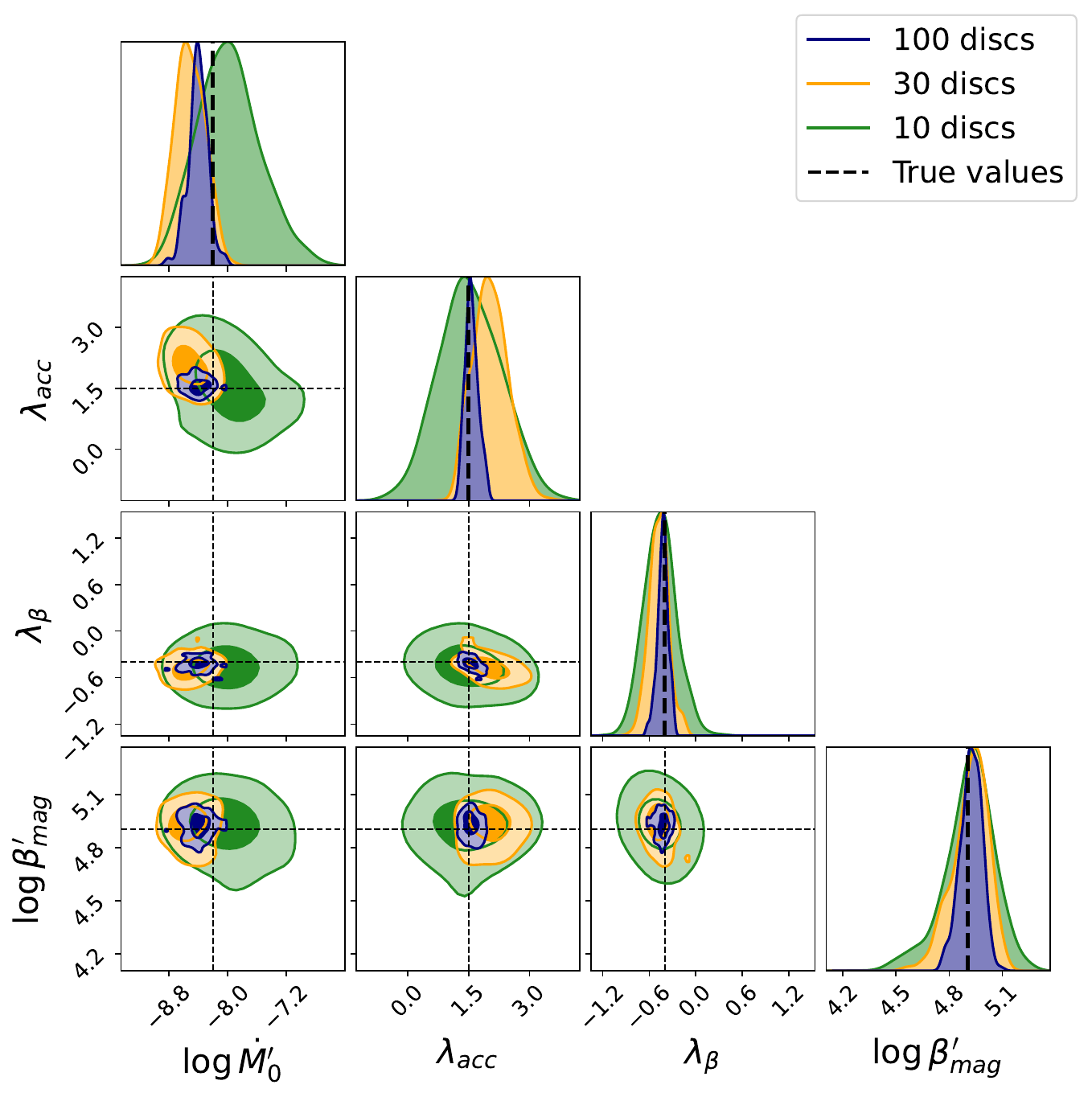}
        \caption{Comparison between the posterior distribution of a subset of the model parameters of a synthetic dataset that evolves according to the MHD-wind model for different subsets of the same synthetic population described it Sect. \ref{test_synthetic_data}. Each subsets contains 100 (blue), 30 (orange), and 10 (green) objects, respectively. Dashed black lines mark the true values, while dark and light hatched regions include the 1$\sigma$ and 2$\sigma$ confidence intervals, respectively.
        }
        \label{fig:test_ndiscs_4_observables}
    \end{figure}

In this section we investigate the impact of the size of the observed sample on the accuracy and precision of the distribution of model parameters.
We considered the synthetic population introduced in Sect. \ref{test_synthetic_data}, generated following the MHD-wind framework, and performed three different \texttt{DEMOS} fits, each based on a subset of 100, 30, and 10 objects, respectively, from such a population.
This test provides a prediction of the precision with which \texttt{DEMOS} can constrain model parameters based on the number of observed discs in populations. A sample of 100 discs corresponds approximately to the size of complete surveys that measure the disc mass from the continuum flux \citep{Manara2023}, while samples of 10 and 30 discs reflect sizes of the new generation of surveys that measure $M$ from CO, namely AGE-PRO \citep{Zhang2025} and DECO (Cleeves et al. in prep.), respectively.
Figure \ref{fig:test_ndiscs_4_observables} compares the posterior distribution of four parameters among the MHD-wind-model set, $\log \dot{M}_0' $, $\lambda_\text{acc}$, $\lambda_\beta,$ and $\log \beta_\text{mag,}'$ obtained in the three scenarios.
As expected, larger samples provide more robust statistics, yielding more accurate and precise constraints on model parameters. Consequently, posterior distributions obtained with ten observed discs (green) are broader than the ones obtained for 30 (orange) and 100 (blue) discs. 
Figure \ref{fig:test_ndiscs} shows the posterior distribution of the full set of parameters.
With the exception of a few parameters such as $\sigma_t$ and $\lambda$, even the ten-disc sample yields informative, though broader, constraints on most parameters, indicating that their impact is significant despite the small sample size.
Therefore, although larger samples result in much more robust constraints, this method is less affected by sample incompleteness compared to others (e.g. \citealt{Alexander2023, Somigliana2024}), because it uses multiple observables. 
This lends credence to previous attempts to model samples of this size, such as \cite{Tabone2025}.
However, if samples exhibit selection biases, which is the case for AGE-PRO and DECO, those biases must also be properly applied to the synthetic populations. 
This can substantially reduce the precision of parameter inference.
Moreover, this tool could be used while designing new observational programmes, as it can estimate the optimal number of sources required constrain the model to a desired precision. 
In addition, \texttt{DEMOS} can provide uncertainties on estimated parameters, which are a combination of the observational uncertainties and on the limited sample size.

\section{Comparison with previous works}
\label{comparison}
After verifying the capability of \texttt{DEMOS} to recover the input parameters of synthetic datasets (see Sect. \ref{tests}), we investigated whether this new population-synthesis tool yields results consistent with the ones obtained by other models in the literature when applied to the same data. 

\cite{Lodato2017} assessed whether the observed $\dot{M}-M$ correlation of the Lupus star-forming region could be reproduced by a purely viscous disc model. 
The authors considered the \cite{Manara2016} sample, where accretion rates are obtained from the VLT-XShooter surveys \citep{Alcala2014, Alcala2017} and the masses are estimated from the sub-millimetre continuum flux \citep{Ansdell2016}.
They generated synthetic populations of discs with initial conditions $M_0$ and $t_\nu$ extracted from log-normal distributions with Monte Carlo methods, with means of $\langle \log M_0 \rangle$ and $\langle \log t_\nu \rangle$ and spreads of $\sigma_{M_0}$ and $\sigma_{t_\nu}$. 
Subsequently, they evolved each disc following the self-similar solution \citep{LyndenBellPringle1974} until an age, $t$, obtained from a log-normal distribution with a mean of $\langle \log t \rangle$ and spread of $\sigma_t$. 
For various combinations of parameters, they assessed which synthetic population most accurately reproduced the observed correlation and spread in the $\log \dot{M} - \log M$ plane.
They considered an uncertainty of 0.1 dex for disc masses and 0.45 dex for accretion rates. 
For $\sigma_{M_0}$ = 0.2 dex, $\sigma_{t_\nu}$ = 1 dex, $\sigma_t$ = 0.3 dex, and $\gamma = 1.5$, where $\gamma$ is the slope of dependence of $\nu$ on $r$, they obtained $\langle \log (M_0 / \text{M}_\odot) \rangle = -2.2$, $\langle \log (t_\nu / \text{yr}) \rangle$ = 5.8 $\pm$ 0.4, and $\langle \log (t/\text{yr})\rangle $ = 5.9 $\pm$ 0.3.
We performed a fit of this sample using a purely viscous framework and considering $M$ and $\dot{M}$ as observables. We set $\langle \log M_0 \rangle $, $\langle \log t_\nu \rangle, $ and $\langle \log t \rangle $ as free parameters, while fixing all other model parameters to the values adopted by \cite{Lodato2017}.
Since the disc fraction is not considered as observable in this fit, the likelihood only consists of the distribution term (only the first term of Eq. \ref{eqn:likelihood_total}).
    \begin{figure}[ht!]
        \centering
        \includegraphics[
                width=\hsize]{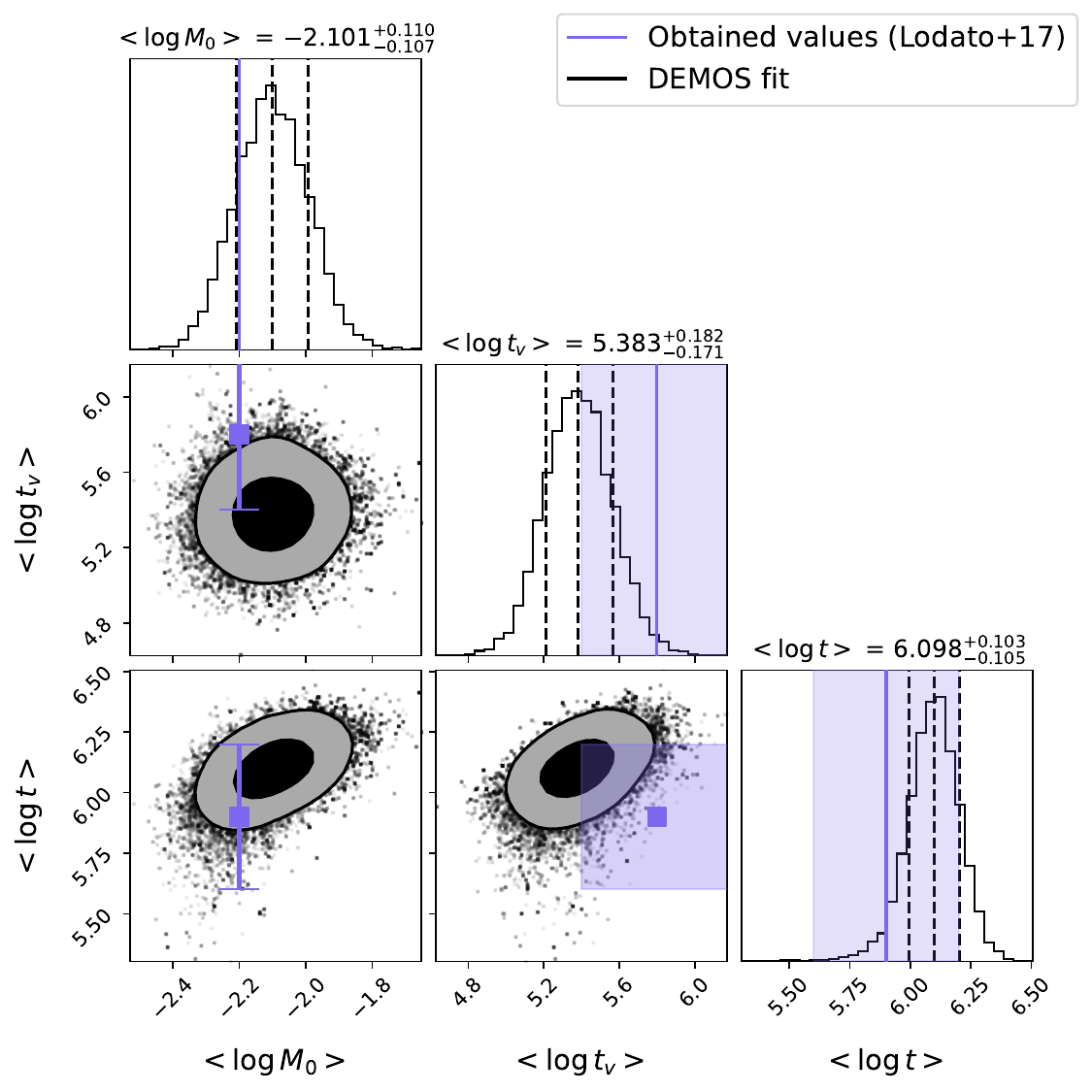}
        \caption{Posterior distribution of the model parameters obtained with \texttt{DEMOS} (black), reproducing the $\log \dot{M} - \log{M}$ plane of the Lupus demographics with a purely viscous model. The distributions are compared with the values inferred by \cite{Lodato2017} (blue). The black and grey shaded areas include the 1$\sigma$ and 2$\sigma$ confidence intervals, respectively.} 
        \label{fig:Lodato+17}
    \end{figure}
Figure \ref{fig:Lodato+17} shows the posterior distribution of the free parameters that we adopted compared to the values obtained by \cite{Lodato2017}. The inferred initial mass and mean age are consistent between \texttt{DEMOS} and \cite{Lodato2017}. In contrast, the mean viscous timescale, $\log t_\nu$, inferred by \texttt{DEMOS} is lower than the one estimated by \cite{Lodato2017} by approximately 1$\sigma$. 
This slight discrepancy should not be associated with a malfunction of either method, since the set of observables used to constrain the models is slightly different.
Indeed, \cite{Lodato2017} only evaluated the compatibility of the synthetic and observed correlation in the $\log \dot{M} - M$ plane, whereas \texttt{DEMOS} analyses the full distribution, thus considering a larger number of observables.

Concerning the MHD-wind model, we compared the predictions of \texttt{DEMOS} with those of \cite{Tabone2022}, which reproduced the Lupus demographics using the analytical prescription of the wind-driven framework \citep{Tabone2022model}.
They considered a more recent dataset of the same observables \citep{Manara2019}, which includes both disc masses and accretion rates.
In the synthetic populations, they drew the initial masses from a log-normal distribution, while they extracted the wind's accretion timescale, $t_\text{acc,0}$, which linearly correlates with the dispersal time, from an exponential distribution compatible with the observed disc fraction \citep{Fedele2010}. 
Considering a mean age of 2 Myr, $\omega = 0.5,$ and the same uncertainties of \cite{Lodato2017}, they obtained as best-fit model of $\langle \log (M_0 / \text{M}_\odot)\rangle  = -2.4$ and $f_M = 1.8$, where $f_M$ is the ejection-to-accretion ratio.
    \begin{figure}[ht!]
        \centering
        \includegraphics[
                width=\hsize]{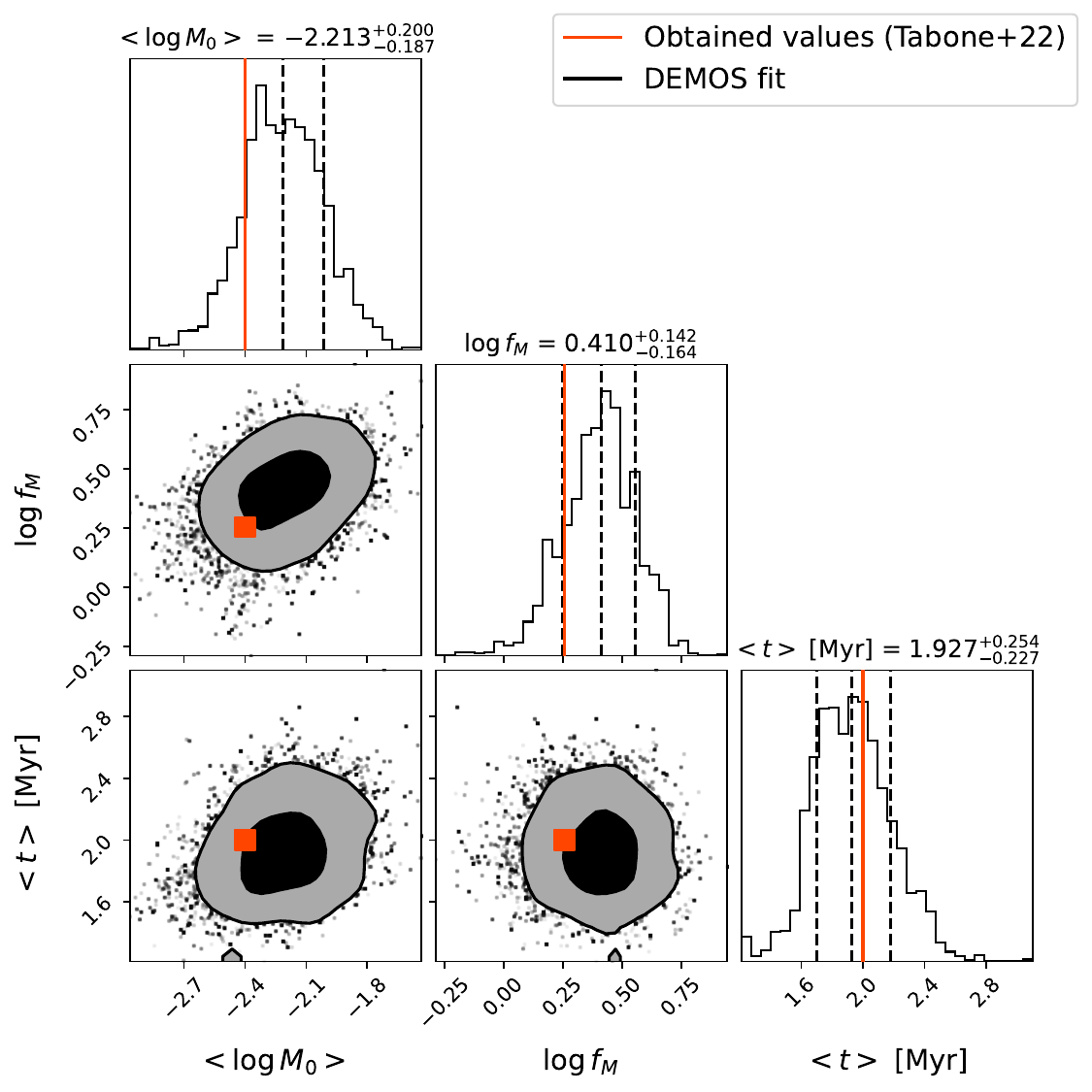}
        \caption{Posterior distribution of the model parameters obtained with \texttt{DEMOS} (black), reproducing the $\log \dot{M} - \log{M}$ plane and the disc fraction of the Lupus demographics with the MHD-wind model. The distributions are compared with the values inferred by \cite{Tabone2022} (orange). The black and grey shaded areas include the 1$\sigma$ and 2$\sigma$ confidence intervals, respectively.}
        \label{fig:Tabone+22}
    \end{figure}
We performed a fit of this sample considering $M$, $\dot{M,}$ and the disc fraction as observables and $\langle \log M_0 \rangle $, $\log f_M,$ and $\langle t \rangle $ [Myr] as free parameters, while setting all other model parameters to the same values of \cite{Tabone2022}.
Figure \ref{fig:Tabone+22} shows the posterior distributions obtained by \texttt{DEMOS} following the MHD-wind model.
The obtained posterior distribution of all parameters is in remarkable agreement with the estimates by \cite{Tabone2022}.
In conclusion, the overall agreement between and these two independent methods
and \texttt{DEMOS} emphasises its robustness as a population-synthesis tool
for inferring model parameters from demographics.

\section{Discussion}
\label{Discussion}
\subsection{The potential of \texttt{DEMOS}}
\label{potential_DEMOS}
The Bayesian inference tool \texttt{DEMOS}, presented in this paper, is a promising framework to infer discs' model parameters from the increasing volume of demographic data \citep{Manara2023}.
Its distinctive ability to exploit the shape of the observed distributions enables us to constrain systematic correlations between disc initial conditions and $M_\star$, which have been shown to be necessary to reproduce the observed correlations \citep{Alexander_Armitage2006, Somigliana2022, Somigliana2024}.
Such analyses are not possible within statistical frameworks that only consider the mean values of observed disc properties \citep{Tabone2025}.
Moreover, although its precision is highest for observed samples of $\sim$100 discs, \texttt{DEMOS} provides meaningful constraints even for smaller samples (see Fig. \ref{fig:test_ndiscs}), thereby extending its applicability to various star-forming regions.
Hence, \texttt{DEMOS} will exploit the available demographics to shed light on the mechanism that drives disc evolution and on the initial conditions that are required to explain observed disc populations. 
These inferred initial conditions can then be compared with theoretical expectations (e.g. \citealt{Bate2018, Lebreuilly2024}) and observations (\citealt{Tychoniec2018, Maury2019, Tobin2020}; see also \citealt{Maury2022} for a review) of Class 0/I objects. 
Furthermore, the model and parameter sets that most accurately reproduce protoplanetary-disc demographics will provide robust, data-driven disc templates. These templates can be used in planet population-synthesis models (e.g. \citealt{Mordasini2009} and \citealt{Emsenhuber2021}, among others) to reproduce exoplanet populations within a framework that remains consistent with the observed disc properties.

In addition to this feature, \texttt{DEMOS} can be used to design observational programmes.
As we showed in Sect. \ref{test_observables}, some observables are necessary to constrain certain sets of parameters. 
For instance, in order to place meaningful constraints on the angular-momentum transport parameter, namely $\alpha$ in the viscous framework and $\alpha_\text{DW}$ in the wind-driven model, even partial measurements of disc radii are needed (cf. Fig \ref{fig:test_observables}; see also Paper II). 
Moreover, the optimal number of sources  to observe can be chosen based on the desired precision in the estimates on model parameters (Sect. \ref{fig:test_ndiscs}).

\subsection{Future improvements}
\label{future_improvements}
The method on which \texttt{DEMOS} is based, described in Sect. \ref{methods}, relies on the comparison between the observed population and synthetic populations of discs, where the stellar mass distribution is generated from typical IMFs.
This comparison is statistically valid only under the assumption that the observed sample is complete.
While this assumption holds for some samples, such as the Lupus and Chamaeleon I regions (except at the lowest stellar masses), included in the \cite{Manara2023} review, it is generally not satisfied.
Indeed, most surveys are affected by either observational or selection biases that can impact the statistical robustness of the comparison.
As a consequence, we plan to introduce suitable biases in \texttt{DEMOS}, in order to address this limitation and extend the analysis to more recent disc surveys (e.g. AGE-PRO, \citealt{Zhang2025}, or DECO, Cleeves et al. in prep.).

In Paper II we present the results of the application of \texttt{DEMOS} to the Lupus star-forming region.
Extending this analysis to more star-forming regions is indeed one of the most compelling next steps, as it would illustrate whether distinct disc populations can be reproduced using a common set of initial conditions.
In addition, due to the larger statistics represented by the entirety of observed demographics, such an extension would place stronger constraints on disc evolution than those obtained for Lupus alone.
In particular, analysing the Upper Scorpius population \citep{Pecaut2016, Luhman&Esplin2020} is particularly stimulating, since it is the most complete and largest sample of an older population (estimated 4-10 Myr).
Therefore, it can provide important constraints on disc populations at a later stage of evolution.
However, Upper Scorpius is located in a mildly irradiated region, where the average far-ultraviolet (FUV) flux, due to massive nearby stars, is approximately 20 $G_0$ \citep{Anania2025}.
\cite{Anania2025AGE-PRO} showed that, under this level of irradiation, the impact of external photoevaporation on global disc properties (particularly disc size and mass) cannot be neglected.
Consequently, we plan to incorporate external photoevaporation into \texttt{DEMOS} in order to reproduce irradiated populations. 
Moreover, we intend to include other environmental effects in future versions of the tool, such as late-stage infall, whose presence has been identified with both direct \citep{Ginski2021, Gupta2023, Speedie2025} and indirect \citep{Huang2024, Winter2024} tracers.

The statistical method on which \texttt{DEMOS} is based relies on the comparison between synthetic and observed populations. This approach is independent of the specific model used to generate the synthetic population, making \texttt{DEMOS} a flexible tool that can be easily extended to various evolutionary models.
In future applications, it will be possible to test disc models that account for the simultaneous presence of turbulence and magnetic winds in discs, as well as solutions in which the angular-momentum transport mechanism and its strength vary spatially across the disc and evolve over time (e.g. \citealt{Gressel2020, Garate2021, Tong2024}).
Furthermore, to enable a proper comparison of synthetic populations to observed demographics inferred from dusty proxies, it will be necessary to implement dust evolution in \texttt{Diskpop} (e.g. \citealt{Pfeil2024}).
The inclusion of binary systems in the tool will further increase the completeness of our estimates about protoplanetary-disc models, due to the paramount role played by binaries in planet formation (see \citealt{Offner2023} for a review).
All these extensions will, however, substantially increase the computational cost of \texttt{DEMOS} fits, because in most cases an analytical solution is not available.
Consequently, we foresee an upgrade of \texttt{DEMOS} incorporating deep-learning techniques, which have already been implemented in the context of protoplanetary discs (e.g. \citealt{Ribas2020, Telkamp2022, Mao2023, Ruzza2024, Ruzza2025}).

\section{Conclusions}
\label{conclusions}

In this paper we presente \texttt{DEMOS}, a Bayesian inference tool designed to infer model parameters from the observed distribution of global disc properties. 
We show how we estimated the PDF of the model by generating synthetic disc populations with \texttt{Diskpop} \citep{Somigliana2024} and calculated the likelihood.
The tool is flexible, allowing for different parameters to be investigated, and it can be extended to include environmental effects in disc evolution, such as late infall or external photoevaporation.
To reduce the computational cost of synthetic populations evolving with the viscous model and internal photoevaporation, we extended the \cite{Ruden2004} solution to all photoevaporative profiles and verified its compatibility with numerical models.
Moreover, we performed multiple tests to verify the correctness of \texttt{DEMOS}.
We used synthetic datasets to assess the impact of different observables on model parameters and how different sample sizes affect the precision of our estimates.
We verified that the results obtained with \texttt{DEMOS} are compatible with those reported in previous works when using the same dataset to constrain the same model parameters.
Our results can be summarised as follows.
\begin{enumerate}
    \item By exploiting the full observed distribution of global star and disc properties (stellar mass, $M_\star$; disc mass, $M$; accretion rate, $\dot{M;}$ and disc radius, $R$) rather than only the mean values, \texttt{DEMOS} can help constrain models with a larger number of free parameters than observables, such as those that assume initial correlations with disc properties and the stellar mass.
    Moreover, \texttt{DEMOS} provides uncertainties on model parameters, reflecting both observational errors and sample size.
    \item Datasets that lack one of the key observables might constrain some free parameters with less accuracy and precision. For instance, to constrain the slopes of the initial correlations it is necessary to include $M_\star$ among the observables. On the other hand, to constrain the angular momentum transport parameter, $\alpha$ or $\alpha_\text{DW}$, measurements of $M$, $\dot{M,}$ and $R$ are required.
    \item Although a sample consisting of $\sim$ 100 discs provides much more precise estimates of the free parameters, even a reduced sample of $\sim$ 10 discs can provide meaningful constraints, albeit with significantly larger uncertainties. 
    Such result makes \texttt{DEMOS} more versatile than other criteria (e.g. \citealt{Alexander2023, Somigliana2024}) that are much more sensitive to sample size. 
    This is because combining different observables gives stronger constraints than evaluating them separately.
    The ability of \texttt{DEMOS} to assess the impact of different observables and sample sizes on parameter estimates and their uncertainties makes the tool suitable for the design of observational programmes.
    \item Extending of the analytical solution by \cite{Ruden2004} to all photoevaporative profiles reproduces the evolution of global disc properties obtained with numerical solutions. Consequently, since it reduces the computational cost by a factor of five, it is particularly suitable for population-synthesis studies. 
    \item When applied to the same dataset, \texttt{DEMOS} produces results consistent with those of \cite{Lodato2017} and \cite{Tabone2022}, emphasising the robustness of the method.
    \end{enumerate}
In conclusion, \texttt{DEMOS} represents a powerful tool to probe disc evolution using observed demographics, since it is designed to optimally exploit the available statistical information.
Its application to observed disc populations will help disentangle the mechanisms driving disc evolution and advance our understandings of planet formation.
In Malanga et al. (2026b) (Paper II), we show its application to the Lupus star-forming region, identifying the evolutionary model and the initial conditions that best reproduce the data.
However, this approach will reach its full potential when applied to various star-forming regions, spanning a range of ages and environments.
This will enable us to assess whether different populations can be reproduced by the same models and input parameters and thus assess the impact of the environment on planet formation.
Moreover, \texttt{DEMOS} can be used to set physically motivated disc templates for planet formation simulations, which can then be employed in planet population-synthesis studies to interpret the rapidly growing body of exoplanet data.

\begin{acknowledgements}
We thank the anonymous referee for their comments
that helped us improve the clarity of the paper. 
We thank Marco Lombardi for the insightful discussion about Bayesian statistics, which improved the clarity and correctness of the analysis.
The authors acknowledge support from the European Union (ERC Starting Grant DiscEvol, project number 101039452), from Fondazione Cariplo, grant No. 2022-1217, from the European Union’s Horizon 2020 research and innovation programme under the Marie Sklodowska-Curie grant agreement No 823823 (Dustbusters RISE project), and from the ERC Synergy Grant “ECOGAL” (project ID 855130). 
This paper was also supported by COST Action CA22133 PLANETS, who funded LM for a PhD school.
GL acknowledges financial contribution from PRIN-MUR 20228JPA3A.
We acknowledge financial support from the project "Space Observations of Proto-planetary Accretion Disc Evolution (SPADE)” in the framework of the "PROgramma di RIcerca Spaziale di base (PRORIS)", financed by the Italian Ministry of University and Research (Ministero dell’Università e della Ricerca; MUR).
Views and opinions expressed are, however, those of the authors only and do not necessarily reflect those of the European Union or the European Research Council. Neither the European Union nor the granting authority can be held responsible for them.

\end{acknowledgements}

\bibliographystyle{aa}

\bibliography{repbib}

\clearpage
\onecolumn
\begin{appendix}
\section{Testing \texttt{DEMOS} with synthetic data}
\label{AppendixA}
\begin{figure*}[ht!]
    \centering
    \includegraphics[width=0.9\linewidth]{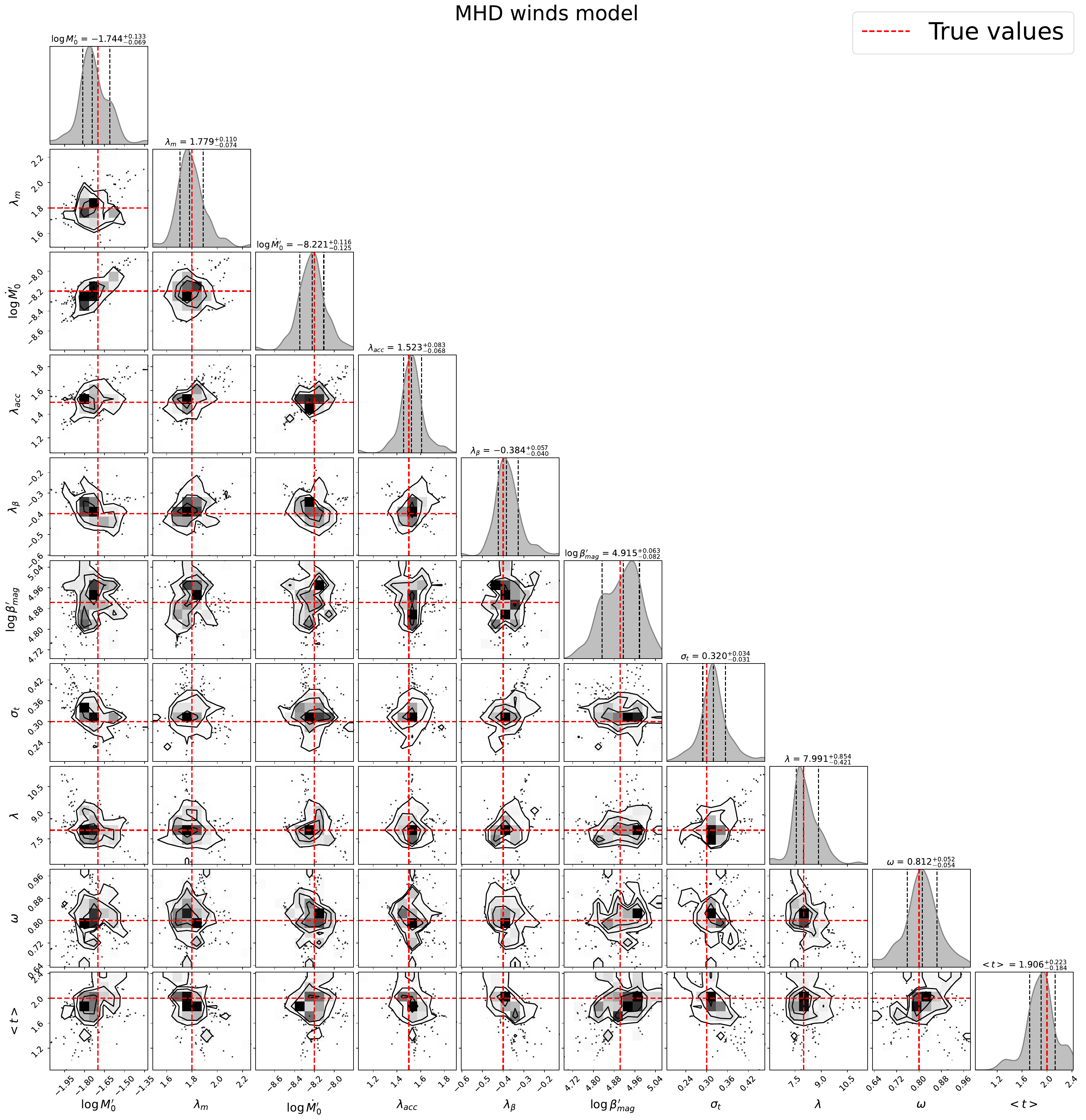} 
     
    \caption{Posterior distribution of the model parameters obtained with \texttt{DEMOS} reproducing a synthetic dataset $D'$ with the MHD winds model. The free parameters $\theta$ are \{$\log (M_0/\text{M}_\odot)', \ \lambda_m, \ \log (\dot{M}_0/\text{M}_\odot \ \text{yr}^{-1})', \ \lambda_\text{acc}, \lambda_\beta, \ \log \beta_\text{mag} ' , \ \sigma_t, \ \lambda , \ \omega , \ \langle  t  \rangle  [\text{Myr}] $\} and the true values (red dashed lines) are \{-1.7, 1.8, -8.2, 1.5, -0.4, 4.9, 0.3, 8.0, 0.8, 2\}. In all cases the true values are compatible with the obtained distributions within 1$\sigma$ intervals.}
    \label{fig:mcmc_mhd_4_variables_corr}
\end{figure*}

\begin{figure*}[ht!]
    \centering
    \includegraphics[width=0.9\linewidth]{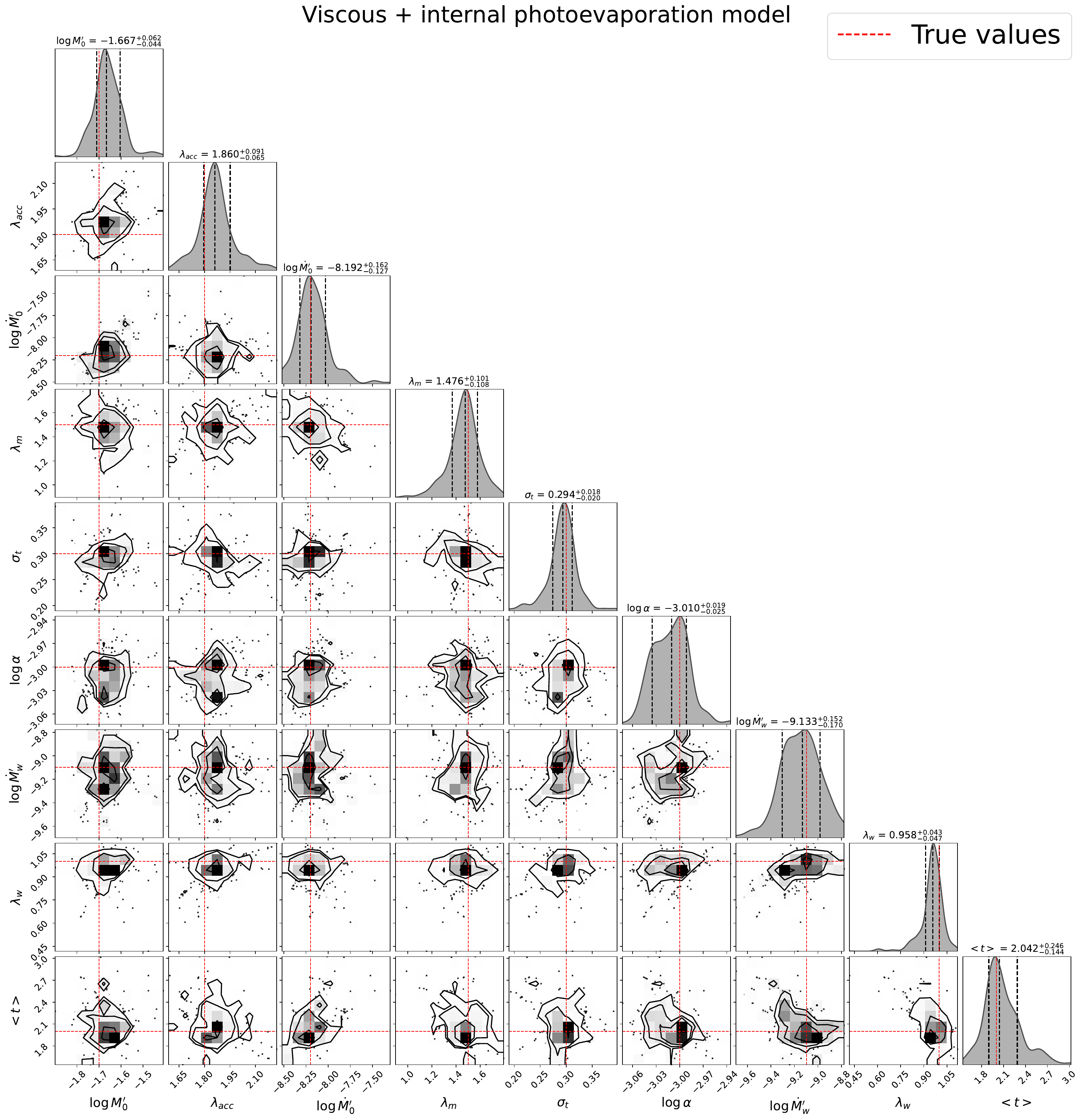} 
     
    \caption{Posterior distribution of the model parameters obtained with \texttt{DEMOS} reproducing a synthetic dataset $D'$ with the viscous-photoevaporative model. The free parameters $\theta$ are \{$\log (M_0/\text{M}_\odot)', \ \lambda_m, \ \log (\dot{M}_0/\text{M}_\odot \ \text{yr}^{-1})', \ \lambda_\text{acc}, \ \sigma_t, \ \alpha, \ \log \dot{M}_\text{w} ' , \ \lambda_\text{w} , \ \langle t \rangle $\} and the true values (red dashed lines) are \{-1.7, 1.8, -8.2, 1.5, 0.3, -3.0, -9.1, 1, 2\}. In all cases the true values are compatible with the obtained distributions within 1$\sigma$ intervals.}
    \label{fig:mcmc_viscous_4_variables_corr}
\end{figure*}

We show the posterior distribution of the model parameters $\boldsymbol{\theta}$, for the MHD-wind model (Fig. \ref{fig:mcmc_mhd_4_variables_corr}) and the viscous model (Fig. \ref{fig:mcmc_viscous_4_variables_corr}), respectively, obtained fitting a synthetic dataset (see Sect. \ref{test_synthetic_data}).

\begin{figure*}[ht!]
    \centering
    \includegraphics[width=0.9\linewidth]{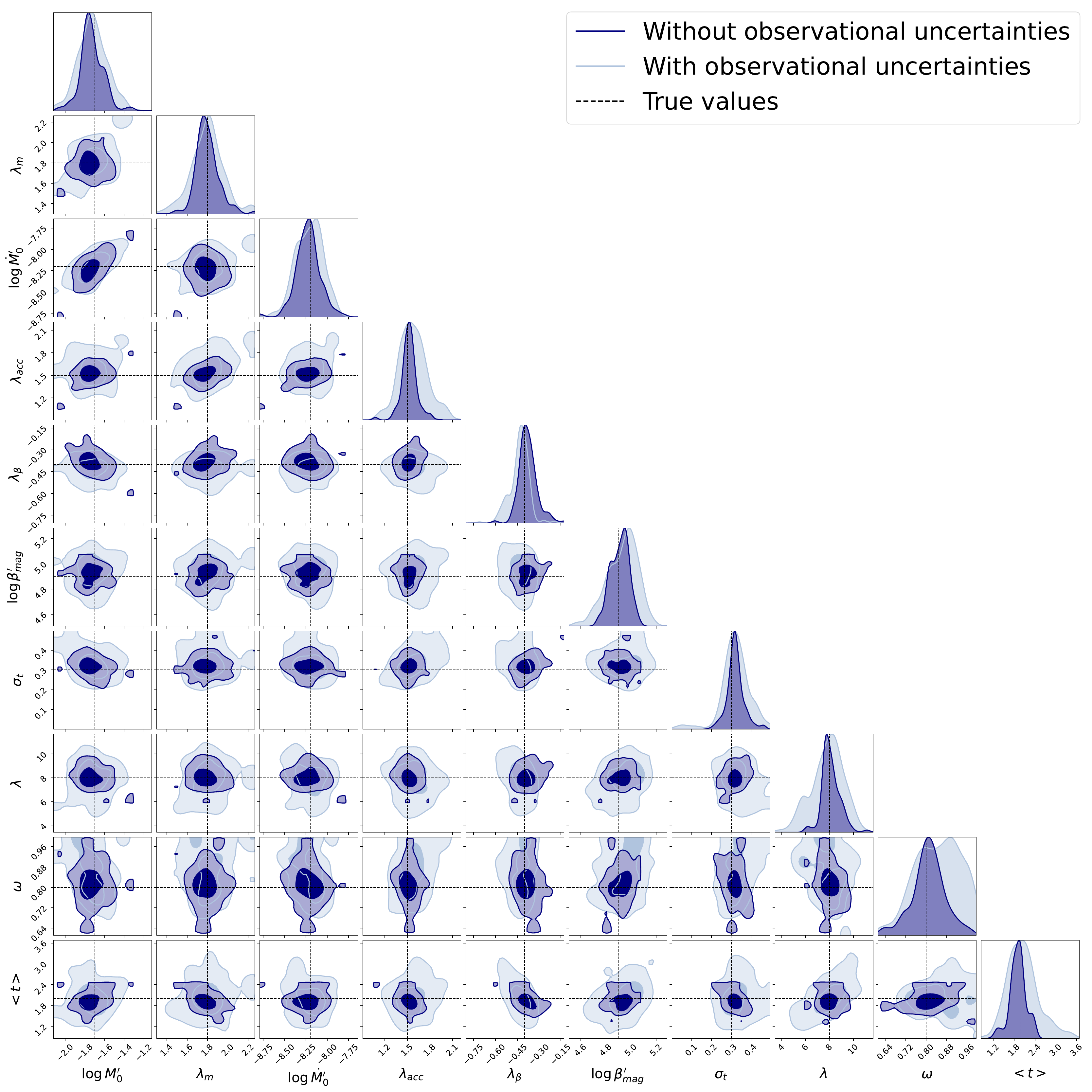} 
     
    \caption{Posterior distribution of the model parameters obtained with \texttt{DEMOS} reproducing the same synthetic dataset of Fig. \ref{fig:mcmc_mhd_4_variables_corr} with (grey) and without (blue) observational uncertainties. Dark and light hatched regions indicate the 1$\sigma$ and 2$\sigma$ confidence intervals, respectively. The dashed black lines mark the true parameter values.}
    \label{fig:mcmc_mhd_4_variables_corr_uncertainties}
\end{figure*}

\begin{figure*}[ht!]
    \centering
    \includegraphics[width=0.9\linewidth]{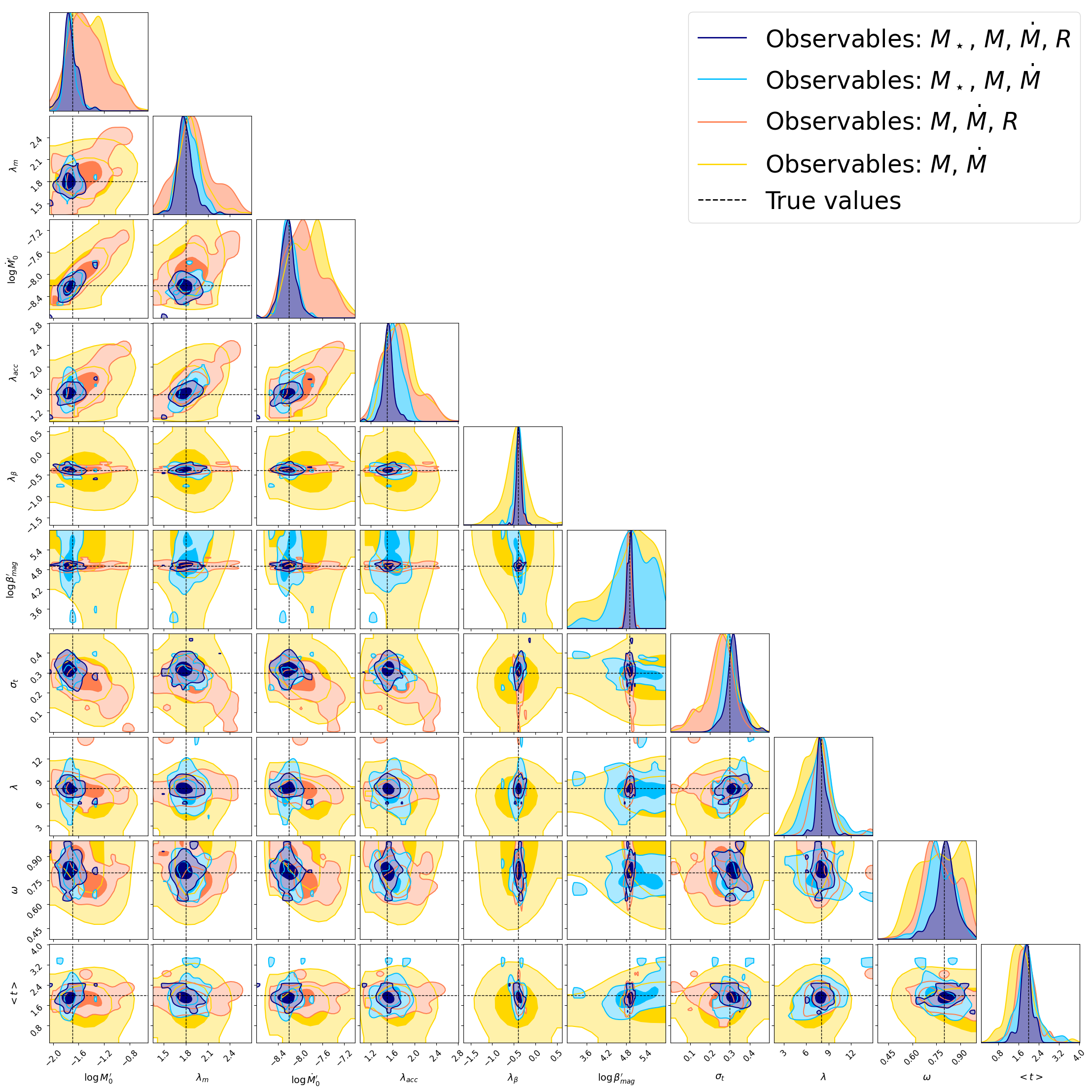} 
     
    \caption{Posterior distributions of all model parameters obtained by fitting the synthetic dataset generated under the MHD-wind model, using different combinations of observables. The observable sets considered are ${M_\star, M, \dot{M}, R}$ (blue), ${M_\star, M, \dot{M}}$ (light blue), ${M, \dot{M}, R}$ (dark orange), and ${M, \dot{M}}$ (yellow). Dark and light hatched regions indicate the 1$\sigma$ and 2$\sigma$ confidence intervals, respectively. The dashed black lines mark the true parameter values.}
    \label{fig:test_observables_total}
\end{figure*}

Figure \ref{fig:mcmc_mhd_4_variables_corr_uncertainties} compares the posterior distributions of MHD winds model parameters, inferred with the same dataset of Fig. \ref{fig:mcmc_mhd_4_variables_corr} (dark blue), obtained without including observational uncertainties, to the case where observational uncertainties are considered (grey).
Figure \ref{fig:test_observables_total} compares posterior distributions of MHD winds model parameters obtained using different subsets of observables of the same dataset.

\begin{figure*}[ht!]
        \centering
        \includegraphics[width=0.9\linewidth]{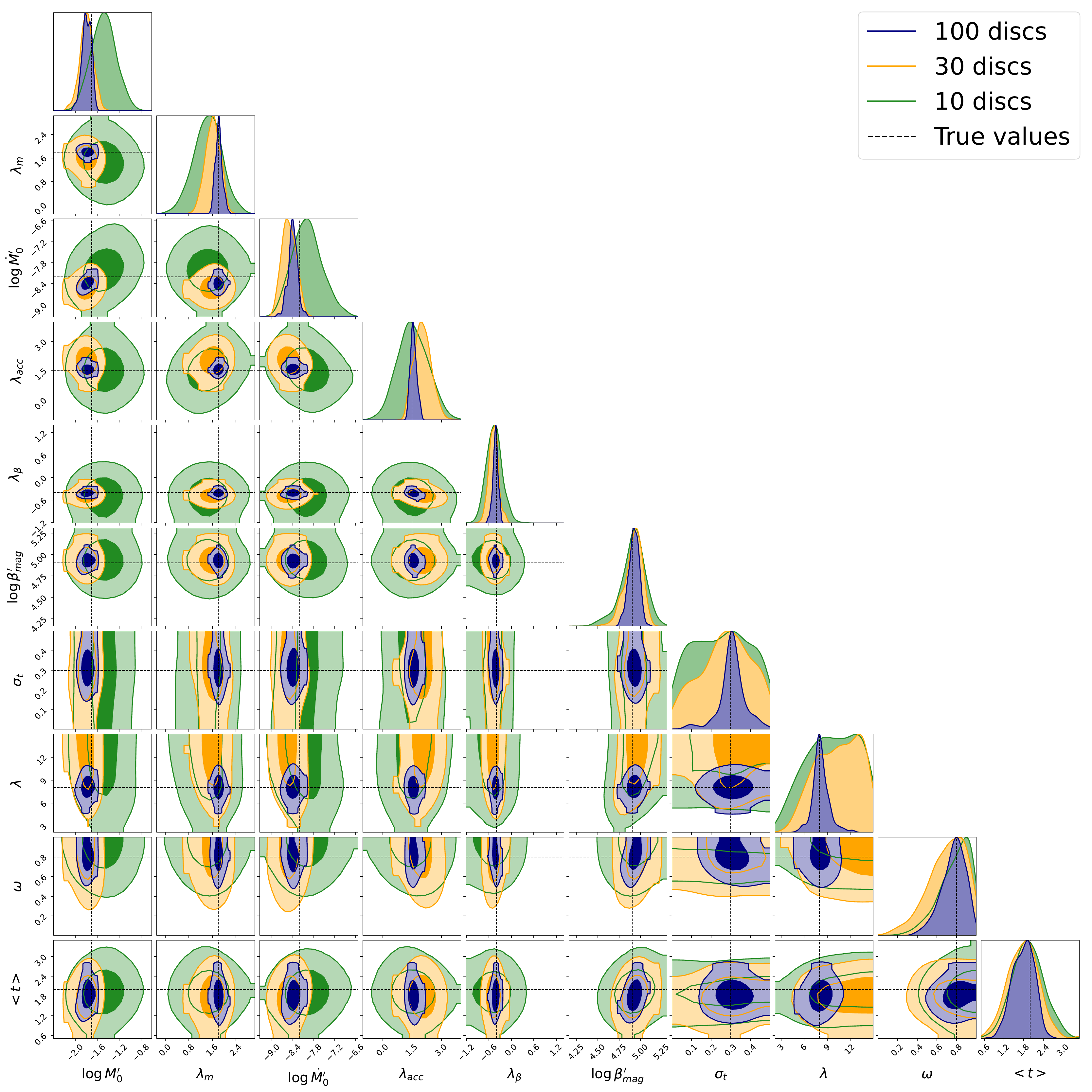} 
         
        \caption{Posterior distribution of the model parameters obtained with \texttt{DEMOS} reproducing the different subsets of the same synthetic population of Fig. \ref{fig:mcmc_mhd_4_variables_corr}. Each subsets contains 100 (blue), 30 (orange) and 10 (green) objects, respectively. Black dashed lines mark the true values, while dark and light hatched regions include the 1$\sigma$ and 2$\sigma$ confidence intervals, respectively.}
        \label{fig:test_ndiscs}
    \end{figure*}

Figure \ref{fig:test_ndiscs} compares posterior distributions of MHD-wind model parameters obtained considering three subsets of 100, 30 and 10 objects, respectively, of the synthetic population defined in Sect. \ref{test_synthetic_data}.
\end{appendix}

\end{document}